\documentclass[iop]{emulateapj}

\makeatletter
\let\@dates\relax
\makeatother

\usepackage{epsfig,amsfonts,amsmath,url,xcolor,wasysym,wrapfig,booktabs,placeins,float,braket,xspace,changes,hyperref}
\usepackage[T1]{fontenc}

\begin{document}

\newcommand{\zmin}{0.65\xspace}
\newcommand{\zmax}{0.9\xspace}
\newcommand{\parzmin}{0.65\xspace}
\newcommand{\parzmax}{0.9\xspace}

\title{Measuring the Luminosity and Virial Black Hole Mass Dependence of Quasar-Galaxy clustering at $ \lowercase{z} \sim 0.8$}
\shorttitle{\uppercase{Luminosity Dependence of Quasar Clustering}}
\shortauthors{\uppercase{Krolewski and Eisenstein}}

\author{
Alex~G.~Krolewski, Daniel~J.~Eisenstein}
\affil{Harvard-Smithsonian Center for Astrophysics, 60 Garden Street, Cambridge, MA 02138, USA}
\email{Corresponding author email: akrolewski@college.harvard.edu}


\begin{abstract} 

We study the dependence of quasar clustering on quasar luminosity
and black hole mass by measuring the angular overdensity of photometrically selected galaxies imaged by WISE about $z\sim 0.8$ quasars
from SDSS.
By measuring the quasar-galaxy cross-correlation function and using
photometrically selected galaxies,
we achieve a higher density of tracer objects and a more sensitive detection of clustering than measurements of the quasar autocorrelation function.
We test models of quasar formation and evolution
by measuring the luminosity dependence of clustering amplitude.  
We find a significant overdensity of WISE galaxies about $z \sim 0.8$ quasars at 0.2--6.4 $h^{-1}$ Mpc in projected comoving separation.
We find no appreciable increase in clustering amplitude with quasar luminosity across a decade in luminosity, and
a power-law fit between luminosity and clustering amplitude gives an exponent of 
$-0.01 \pm 0.06$ (1 $\sigma$ errorbar).
We also fail to find a significant relationship between clustering amplitude and black hole mass,
although our dynamic range in true mass is suppressed due to the large uncertainties in virial black hole mass estimates.
Our results indicate that a 
small range in host dark matter halo mass maps to a large range in quasar luminosity.
\end{abstract}

\section{Background}
\label{sec:intro}

Direct observations have recently confirmed that nearly every nearby elliptical galaxy
or spiral bulge hosts a central supermassive black hole \citep{kr+95}.  
Further studies have shown that black hole properties are tightly
related to the macroscopic properties of the host galaxy.  Black hole mass is closely tied to both the
stellar velocity dispersion \citep{geb+00,fm+00} and the mass of the associated spheroid \citep[e.g. spiral bulge or elliptical galaxy;][]{mh+03, mag+98}.
Both the velocity dispersion and spheroid mass are related to the mass of the host dark matter halo \citep{ferr+02}.
These relationships provide evidence for a link between black hole and host galaxy evolution \citep{sr+98,hopun+06}.

Local supermassive black holes are relics of past quasar activity \citep{solt+82}.
Quasars
are an important stage in the coevolution of a black hole and its host galaxy.
Quasar activity is associated with massive, disruptive flows of gas into the center of the galaxy,
and simulations predict that quasar activity suppresses star formation,
leading to a link between black hole mass and host spheroid mass
 \citep{dim+05}.

However, the details of quasar formation and evolution are not fully known.  Many different models have been proposed to explain quasar activity, 
ranging in sophistication from numerical simulations of quasar
activity tracking gas hydrodynamics \citep{hop+08,deg+11,thack+09,cha+12}
to semi-analytical methods tracking only dark matter halos \citep{kh+00,kh+02,bon+09,fan+12}
to scaling relations with no preferences for the underlying physics \citep{cw+13,lidz+06,crot+09,shen_mod+09,bs+10}.
A brief, comprehensive, and useful overview of various quasar models
can be found in Appendix B of \citet{whi+12}.
In this paper, we consider the differences between low and high luminosity quasars,
and between low and high mass black holes.  Specifically, we study the relationships between luminosity
and host halo mass and between black hole mass and host halo mass by measuring quasar clustering.

Quasar clustering is a valuable tool for exploring quasar properties and evolution.
Quasar clustering is often measured using the linear quasar bias $b_{Q}$, defined by
\begin{equation}
\xi_{Q} = b_{Q}^{2} \xi_{\textrm{matter}}
\label{eqn:bias}
\end{equation}
where $\xi_{Q}$ is the quasar autocorrelation function and $\xi_{\textrm{matter}}$ is the correlation function of the linear-regime matter field.  The bias is
then converted to the host halo mass using analytic formulas tested against $N$-body simulations \citep{ck+89,mow+96,she+01},
which predict that at fixed redshift the halo mass is a monotonically increasing function of bias.
Measurements of the luminosity dependence of quasar clustering allow estimation of the slope and scatter of the relationship
between luminosity and host halo mass.
The halo mass-luminosity relationship
arises from the combination of a black hole mass-luminosity relationship and a
halo mass-black hole mass relationship.  By also measuring the black hole mass dependence of clustering,
we can separate the contributions of the black hole mass-luminosity relationship from the halo mass-black hole mass relationship.

The quasar lifetime can be determined from measurements of the quasar bias \citep{ck+89,hh+01,mw+01}.
If quasars are long-lived, then most of the quasars in the universe are actively accreting and quasars are rare, whereas if quasars are short-lived,
there are many unobserved dormant quasars for each observed active quasar and quasars are common.  Since high mass, highly biased halos are rare \citep{ps+74}
a highly biased quasar population implies a long lifetime.  The conversion between bias and lifetime assumes
that luminosity is a monotonic function of halo mass with no scatter.  If there is scatter in the luminosity-halo mass relationship,
the lifetime is greater than for the no-scatter case \citep{mw+01}.  Thus, measurements of the luminosity dependence of quasar clustering
allow for joint determination of the quasar lifetime and the scatter in the luminosity-halo mass relationship \citep{shank+10}.
The analysis of \citet{mw+01} and \citet{hh+01} assumed that the quasar lightcurve is a step function, either ``on'' or ``off'' at any given time.
More sophisticated models of the quasar lightcurve have been proposed \citep[e.g.][]{kh+02,lidz+06}, leading to predictions of luminosity
or redshift-dependent lifetimes.  In principle, measurements of the luminosity dependence of clustering can help distinguish between
different models of the quasar lightcurve.

Large surveys such as the Sloan Digital Sky Survey \citep{yor+00} and the 2dF QSO Redshift Survey \citep{cro+04}
provide the large samples necessary to study the luminosity dependence of quasar clustering.
Early studies measured the luminosity dependence of the real-space and redshift-space autocorrelation
function using spectroscopically selected quasars:  \citet{cro+05} found no luminosity dependence of the redshift-space clustering strength
of 2dF quasars at $0.5 < z < 2.5$, and \citet{por+06} found no luminosity dependence on real-space clustering strength for 2dF quasars at
$0.8 < z < 1.3$ and only marginal evidence at $1.3 < z < 2.1$.  More recently, \citet{sha+11} found no luminosity dependence on clustering strength at $z = 1.4$
using quasar autocorrelation measurements from 2QZ, 2SLAQ, and SDSS, and \citet{whi+12} found no luminosity dependence on redshift-space clustering
strength for SDSS quasars at $z = 2.4$.
Due to the low space density of spectroscopically detected quasars, these studies generally suffer from large error bars.
One way to
obtain smaller error bars is to measure the autocorrelation function of photometrically selected quasars.  \citet{my+07} measured the autocorrelation of 300,000
photometric quasars from SDSS and found no luminosity dependence of quasar bias at photometric redshifts $z = 0.85$, $z = 1.44$,
and $z = 1.92$.

Another approach is to measure the cross-correlation of galaxies about quasars.  Since galaxies have a much higher spatial density than quasars,
cross-correlations can achieve lower error bars than autocorrelations \citep[as suggested by][]{kh+02}.  Early quasar-galaxy cross-correlations used small samples and thus
had large error bars.  \citet{ade+05}, measuring the cross-correlation of quasars at 
$2 < z < 3$ with Lyman-break galaxies, found no luminosity dependence across 4 decades in quasar luminosity,
but urged caution in interpretation of their result due to large error bars.  Similarly, \citet{c+07} also found no luminosity dependence on
quasar-galaxy clustering at $z \approx 1$, but again with relatively large error bars. 
More recent studies found little evidence for luminosity-dependent clustering with larger samples and smaller error bars \citep{daa+08,shen+09,moun+09,shir+11,shen+13},
although \citet{shen+09} did find an increase in bias for the most luminous 10\% of quasars.

Some studies used photometrically detected galaxies rather than
spectroscopic galaxies to increase the size of the tracer population and further decrease their statistical errors.
\citet{pad+09} found no luminosity evolution in the cross-correlation between SDSS DR5 quasars
and luminous red galaxies with photometric redshifts at $0.2 < z < 0.6$. \citet{zh+13} found no luminosity dependence in the cross-correlation between
DR5 quasars and photometrically detected galaxies from SDSS Stripe 82.

While most studies focused on the optical luminosity dependence of quasar clustering, other studies measured the luminosity dependence at different wavelengths.
\citet{hick+09} found significant variation in clustering strength
between radio, X-ray, and infrared selected AGNs.  Using X-ray selected quasars from ROSAT,
\citet{krum+12} found a 2 $\sigma$ dependence of clustering strength on X-ray luminosity, and
attributed the discrepancy between measurements of the optical luminosity dependence and measurements of the X-ray luminosity
dependence to the much larger dynamic range in X-ray luminosity.

Fewer studies have attempted to measure black hole mass dependent clustering due to the much higher uncertainties of black hole mass estimates.
Black hole masses can be estimated from single-epoch spectroscopy using continuum luminosity and emission line width \citep{vp+06} with uncertainties
up to 0.5 decades for individual objects \citep{shen_mass+13}.
Throughout this paper, we refer to these estimates as ``virial black hole masses'' to emphasize the large associated uncertainties.

Using the same sample as \citet{cro+05}, \citet{fine+06} constructed composite spectra for each of 10 bins in redshift and measured the virial mass for each
composite spectrum.
They found a $M_{\textrm{BH}}-M_{\textrm{DMH}}$ relationship in good agreement with the models of \citet{ferr+02}, although 
with substantial uncertainties in the slope and zeropoint of the relation due to both the small dynamic
range in virial mass and the Malmquist bias arising from the 2QZ flux limit.
Other studies measured the evolution of clustering with virial mass at fixed redshift: \citet{shen+09}
found no dependence of quasar clustering on virial mass using two bins in viral mass,
and \citet{zh+13} found a 1--2 $\sigma$ difference in clustering strength between two bins in virial mass.
However, these studies were hampered by a small number of bins and a small dynamic range in virial mass,
implying an even smaller range in true mass due to the uncertainty in the virial masses
 \citep{shen+09}.
\citet{komi+13} studied the virial mass dependence of AGN clustering across a wide range of redshifts ($0.1 < z < 1.0$) and luminosities,
combining samples from SDSS DR4 and DR7 to measure clustering across 2 decades in virial mass.
They found a significant trend of increasing clustering strength with increasing virial hole mass across
4 bins in virial mass, with a 2--3 $\sigma$ difference in clustering strength between the highest and lowest mass groups.

In this study, we measure the angular overdensity of photometric galaxies about spectroscopic quasars to obtain as numerous
a tracer population as possible, minimizing statistical errors.  We measure the luminosity and black hole mass dependence of the quasar-galaxy
clustering amplitude at $\zmin < z < \zmax$,
using a galaxy sample drawn from the Wide-Field Infrared Survey Explorer \citep{wri+10} and quasar samples from the SDSS DR7 \citep{shen+11} and DR10
quasar catalogs \citep{par+13}.  Because the WISE galaxies lack spectroscopic redshifts, we measure the angular correlation
using only the galaxy and quasar positions on the sky.  Since emission from galaxies at $z \sim 0.8$ peaks in the near-infrared, we expect
that WISE selection will maximize the size of our tracer population.

In this paper we begin by discussing the quasar and galaxy samples and their selection criteria (Section~\ref{sec:sample}).  Then we present the angular
clustering measurement (Section~\ref{sec:angular_overdensity}) and find the dependence upon luminosity and virial black hole mass (Section~\ref{sec:results}).  Finally we discuss the
significance of these results and compare them to previous studies (Section~\ref{sec:disc}).  Throughout this paper, we use a $\Lambda$CDM
cosmology with $h = 0.7$, $\Omega_{\Lambda}$ = 0.7 and $\Omega_{m}$ = 0.3 \citep{sperg+03}, matching the cosmology used to construct the DR7 and
DR10 quasar catalogs.  When computing the power spectrum of the linear-regime matter field, we use the transfer function of \citet{eh+98} and $\Omega_{b} = 0.044$, $T_{\textrm{CMB}} = 2.726$ K, $N_{\nu} = 3$,
and $n_{s} = 0.93$ \citep{sperg+03}.
All distances are measured in comoving $h^{-1}$ Mpc.

\section{Quasar and Galaxy Sample Selection}
\label{sec:sample}

We use a color-selected sample of $z \gtrsim 0.6$ galaxies from WISE and $0.65 < z < 0.9$ quasars from
the SDSS DR7 and DR10 quasar catalogs to measure quasar-galaxy clustering.  The galaxy sample was chosen to maximize purity rather than
completeness, allowing us to obtain a high signal-to-noise measurement of quasar-galaxy clustering.
Galaxy colors are measured by comparing SDSS and WISE imaging.

\subsection{Galaxy selection: SDSS and WISE}
\label{sec:gal_selection}

The Wide-Field Infrared Survey Explorer is a satellite that mapped the entire sky at 3.4, 4.6, 12 and 22 $\mu$m
at sensitivities corresponding to Vega magnitudes of 16.5, 15.5, 11.2, and 7.9 in unconfused
regions of the sky \citep{wri+10}.
WISE collected at least 12 exposures at each point on the sky, with coverage depth increasing
rapidly towards the Ecliptic poles due to the WISE scan strategy. 
We select objects from the All-Sky Data Release Source Catalog, which contains 563 million
objects that are not flagged as an image artifact and have both SNR > 5 in at least one band and detections in at least 5 single-band exposures.
The All-Sky Data Release uses imaging taken from January to August 2010,
and extensive documentation can be found at \url{http://wise2.ipac.caltech.edu/docs/release/allsky/}.
We use imaging from the $W1$ band at 3.4 $\mu$m, which has angular resolution of 6" \citep{wri+10}.
The WISE limiting magnitude varies across the sky, most notably due to increased source confusion at low Galactic latitude\footnote{See Figure 9 at \url{http://wise2.ipac.caltech.edu/docs/release/allsky/expsup/sec2\_2.html}.}.
We only consider objects with Galactic latitude $b > 25^{\circ}$; at these latitudes, the $W1$ magnitude limit
is approximately constant.  WISE photometry is given in Vega magnitudes, uncorrected for Galactic dust extinction,
which is negligible at 3.4 $\mu$m.  For sources with $15.5 < W1 < 16$, similar in brightness
to those selected in our galaxy catalog, the astrometric accuracy of WISE is $\approx$ 0.4", as measured by the RMS of the WISE-2MASS
positional difference\footnote{\url{http://wise2.ipac.caltech.edu/docs/release/allsky/expsup/sec6\_4.html}.}.

The Sloan Digital Sky Survey imaged an area of 14555 deg$^{2}$ in 5 filters \citep[$ugriz$;][]{fuku+96}, mostly at high Galactic latitude, using a 2.5 m wide-field telescope
at Apache Point Observatory in New Mexico \citep{gunn+06} and a camera with 30 2048 by 2048 CCDs \citep{gunn+98}.
Imaging data was collected only under photometric conditions \citep{hogg+01} and
photometry was calibrated to 1\% accuracy \citep{smi+02,ivez+04,tuck+06,pad+08} with negligible spatial variation in photometric calibration \citep{fuku+04}.
All SDSS magnitudes cited in this paper are corrected for Galactic extinction using the Schlegel-Finkbeiner-Davis dust map \citep{sch+98}
and are reported as asinh magnitudes \citep{lup+99} in the AB system \citep{fuku+96}.
SDSS photometry is 95\% complete at $r = 22$ \citep{edr+02}.  
Astrometry is typically accurate to 0.1" \citep{pier+03}, although the photometric data used in this paper \citep[DR8;][]{dr8+11} contain an astrometric calibration error that
causes a shift of 0.24" north and 0.05" west over a large region covering most of the survey with declination $ > 41^{\circ}$ \citep{dr8bug+11}.
However, the effect of the astrometry error on our work is negligible.

The SDSS photometric pipeline separates extended sources from point sources based on the difference between the PSF magnitude
and a composite model magnitude consisting of a linear combination of de Vaucouleurs and exponential light profiles \citep{edr+02}.
Many galaxies at
$z \sim 0.8$ are unresolved, particularly in regions of poor seeing.  To 
eliminate seeing-dependent variations in galaxy sample density, our sample contains
point sources as well as extended sources in SDSS imaging.  For both point
sources and extended sources, we use the best-fit
exponential or de Vaucouleurs model magnitudes \citep{edr+02}.
We select our objects using the catalogs from the DR8 data sweeps, which contain point sources with at least one extinction-corrected PSF magnitude
less than [22.5, 22.5, 22.5, 22, 21.5] ($ugriz$) and extended sources with at least one extinction-corrected model magnitude less than [21, 22, 22, 20.5, 20.1] ($ugriz$) \citep{blant+05}.

\begin{figure}[H]
\centerline{\psfig{file=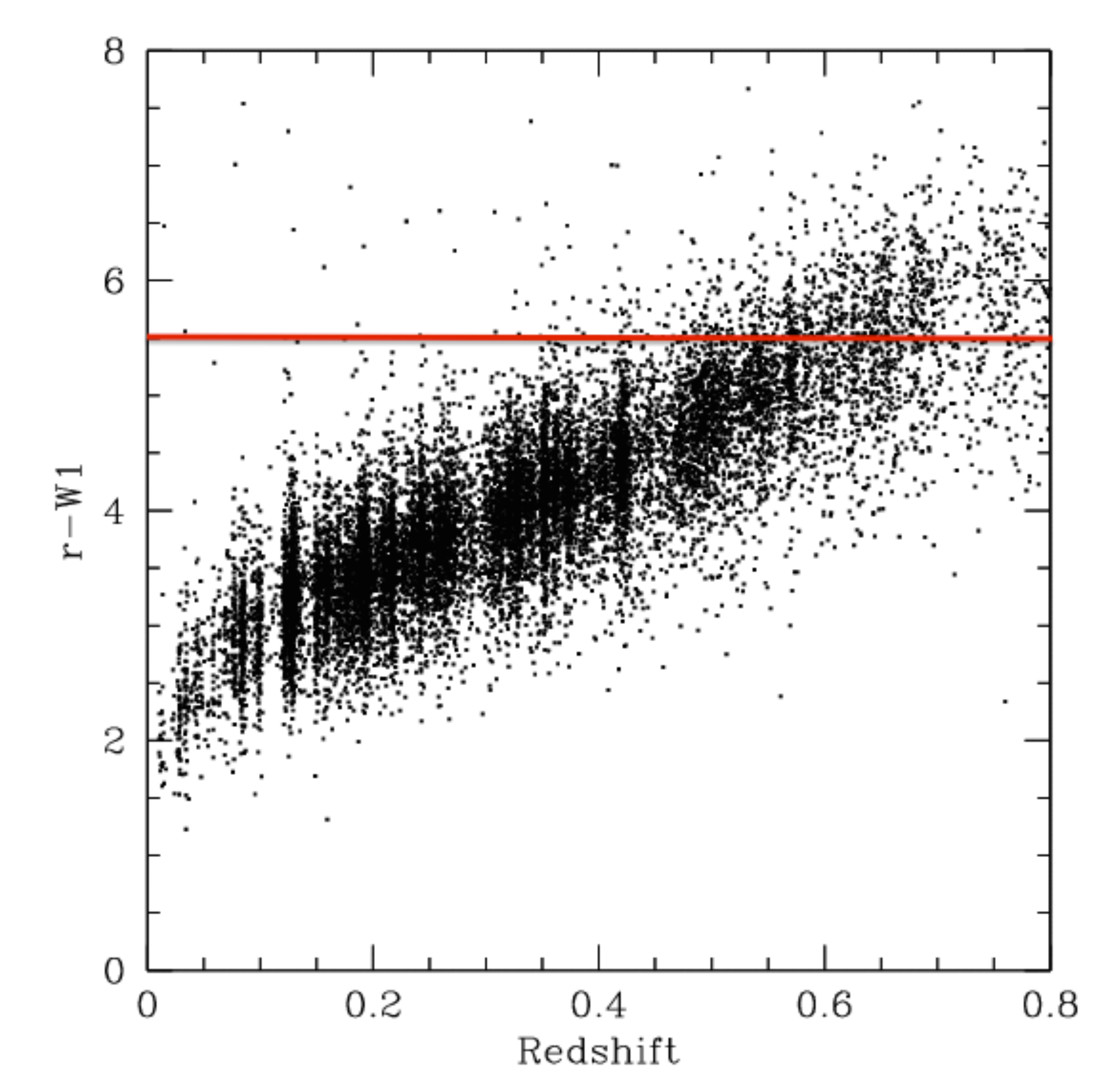,width=8 cm,angle=0}}
\caption[]{\small Redshift vs. extinction-corrected color for a sample of galaxies from the AGES survey.  
Note that $r$ is reported in AB magnitudes while $W1$ is reported in Vega magnitudes.
All points above the red line would be included
by our color cut.  Since most of these galaxies have $z > \zmin$, our color cut removes
many contaminating $z < 0.6$ galaxies.  Almost all of
the severe deviations from the linear trend are objects with spurious colors and should be ignored.}  
\label{fig:AGES}
\end{figure}

We use a color cut to ensure that our sample is composed primarily of $z \gtrsim 0.6$ galaxies.
In order to choose an appropriate color cut, we match WISE and SDSS photometry to spectroscopic redshifts
for 20,000 galaxies measured by the AGN and Galaxy Evolution Survey (AGES) \citep{koch+12}.
To calculate WISE-SDSS colors, we use $r$ magnitudes in the AB system
and $W1$ magnitudes in the Vega system.
Color cuts based on WISE bands  alone are inadequate,
so instead we select all galaxies with $r - W1 > 5.5$.
This is a conservative color cut that minimizes the number of stars and low-redshift galaxies
in our sample, reducing statistical uncertainty in our clustering measurement.
As Figure~\ref{fig:AGES} shows, the vast majority of the galaxies satisfying our color cut have $z > \zmin$.
However, many galaxies with $z > \zmin$ have $r - W1 < 5.5$, particularly blue
galaxies with $0.6 < z < 0.8$.  In this study, we assume that the luminosity and virial mass dependence of
clustering is the same for red galaxies as for blue galaxies.

We begin by choosing WISE detections with \hbox{$14 < W1 < 16.5$}, $W1$ SNR $> 5$ and Galactic latitude $b > 25^{\circ}$.
Less than 0.1\% of the BOSS luminous red galaxies
with $0.65 < z < 0.9$ have $W1 < 14$\footnote{These are mostly CMASS galaxies; see \citet{whi+11} for the sample selection criteria.}, so cutting objects with $W1 < 14$
reduces stellar contamination without removing $z \gtrsim 0.6$ galaxies.
We also remove potentially variable,
saturated, or contaminated detections and detections that were deblended more than once
(W1SAT > 0, W1CC\_MAP > 0, VAR\_FLG > 5, and NB > 2)\footnote{See \url{http://wise2.ipac.caltech.edu/docs/release/allsky/expsup/sec2\_2a.html} for definitions.}.
Since nearly every $z \gtrsim 0.6$ galaxy is unresolved in WISE imaging, we remove detections that are extended sources in WISE
(EXT\_FLG > 0).  Last, we remove detections with more than 10\% of observations contaminated by scattered moonlight (MOON\_LEV > 1).  Since the detections with
moonlight contamination are highly clustered due to the WISE moon avoidance maneuvers\footnote{See \url{http://wise2.ipac.caltech.edu/docs/release/prelim/expsup/sec3_4a.html}.},
we exclude
all WISE detections in regions with high moon contamination.

\begin{figure}[H]
\begin{tabular}{cc}
\psfig{file=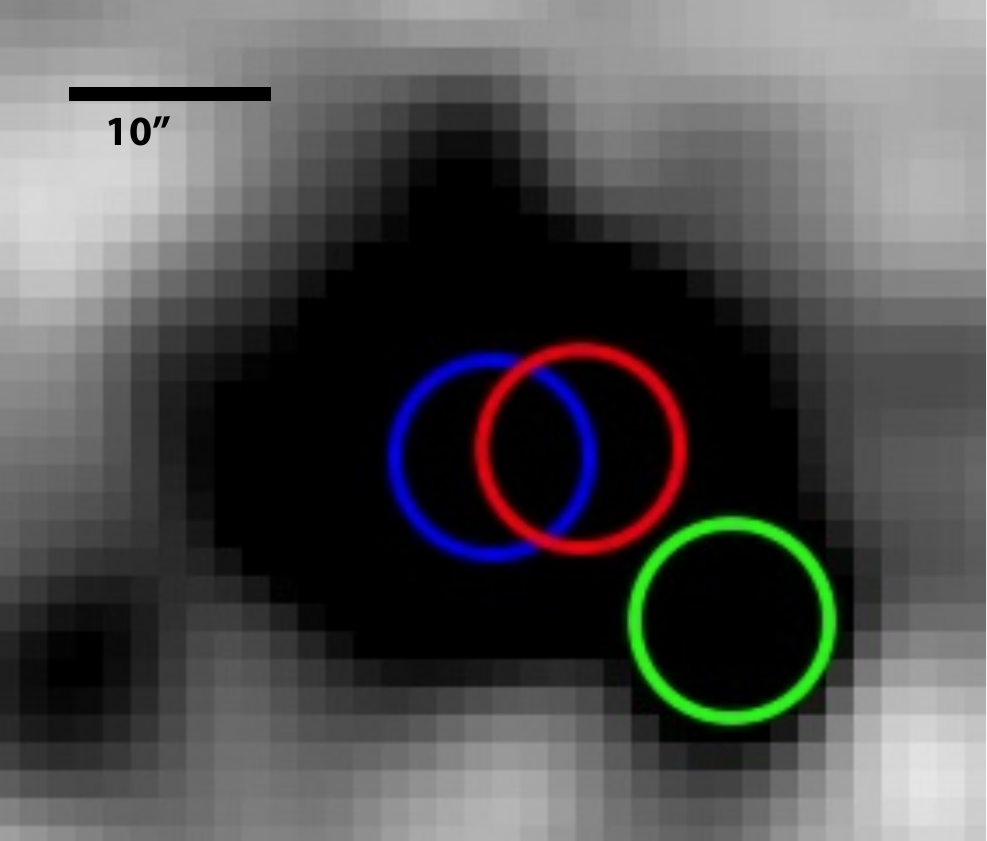,height=3.8 cm,angle=0} &
\psfig{file=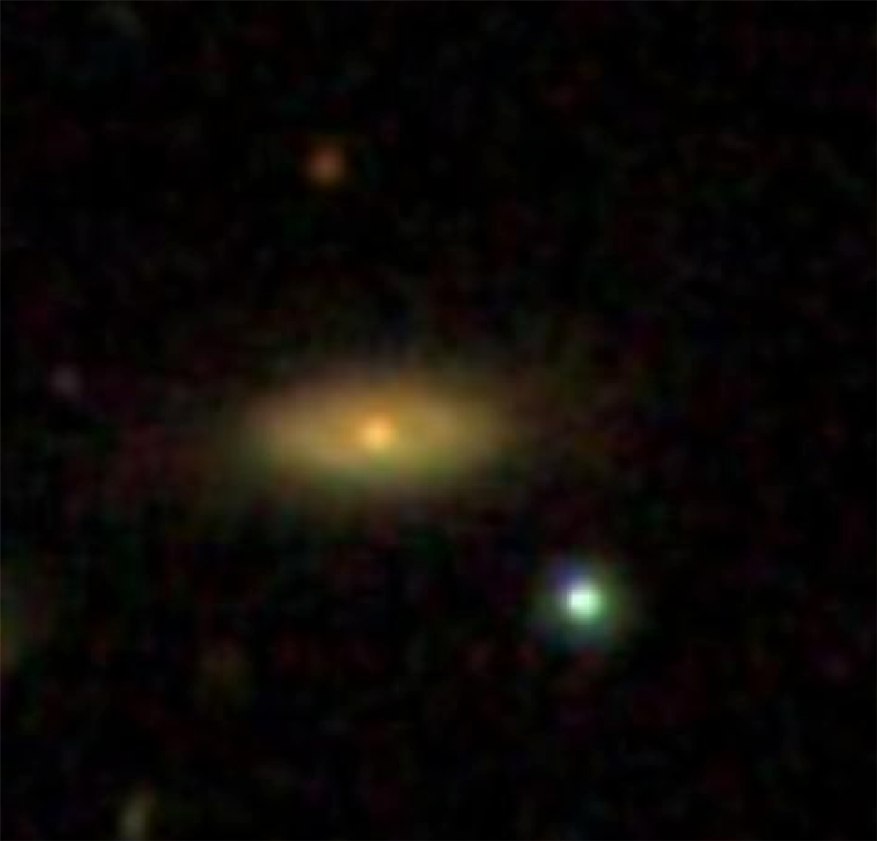,height=3.8 cm,angle=0}
\end{tabular}
\psfig{file=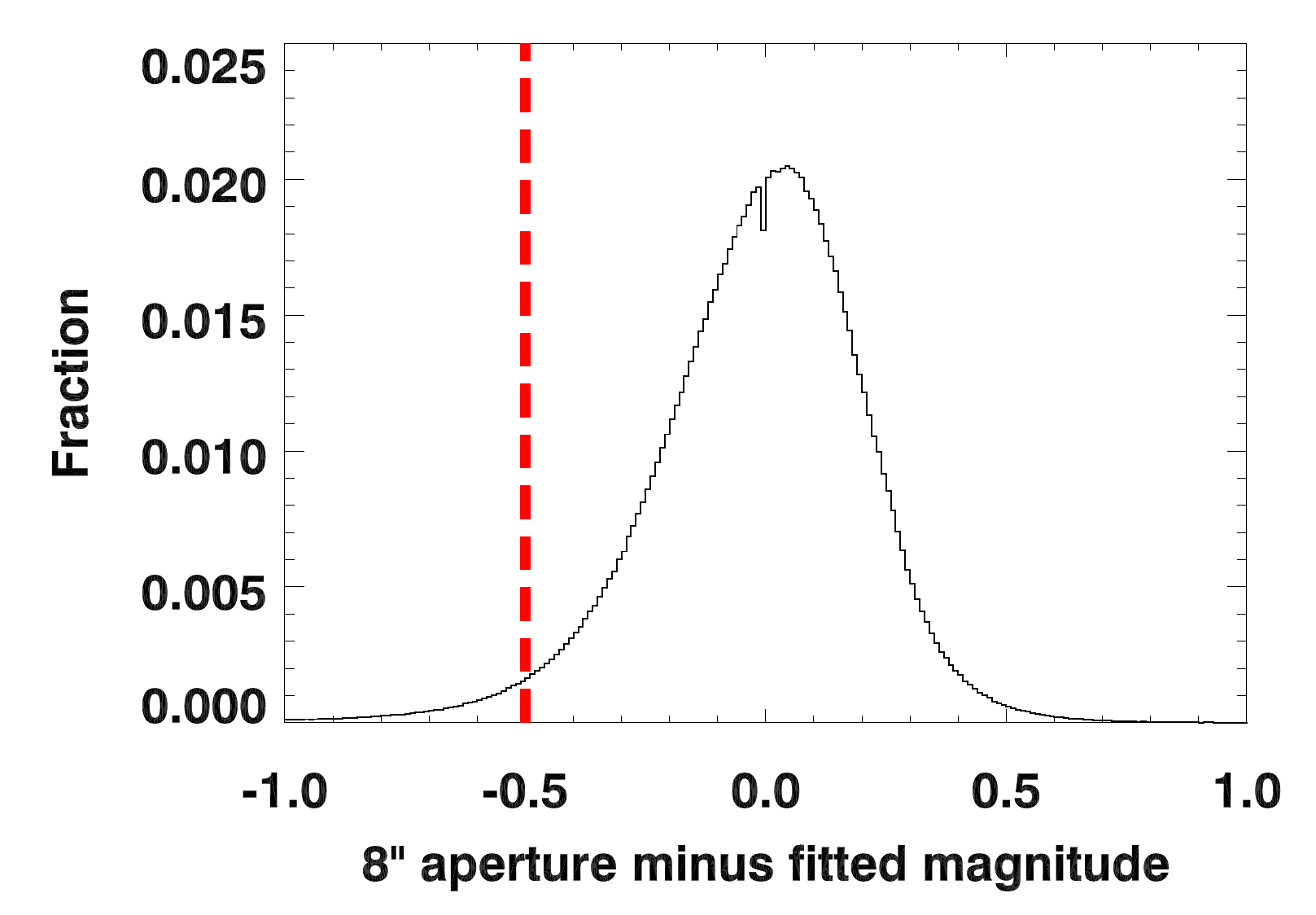,width = 8 cm,clip=}
\caption[]{\small \textit{Top}: example of a WISE blending problem in WISE (left) and SDSS (right),
located at $22^{\textrm{h}} 52^{\textrm{m}} 9^{\textrm{s}}.02, 1^{\circ} 38 ' 6.4 ''$.  The two images cover the same region of the sky.
The colored circles on the WISE image are centered on WISE detections; the red circle lacks an SDSS match.
It is clear that the red and blue circles are both associated with the
bright yellow galaxy at the center of the SDSS image and the red circle is merely an artifact of the WISE deblending process.
\textit{Bottom}: Distribution of $W1$ aperture minus fitted magnitude ($\textrm{A} - \textrm{F}$) for WISE detections included in our sample.  The tail on the left results from 
deblended sources such as those in the top images, which have an extended flux profile despite being classified as point sources.  Because these could
be contaminating low-redshift objects, we removed all detections with $\textrm{A} - \textrm{F} < -0.5$.}  
\label{fig:blending}
\end{figure}

Visual inspection of WISE detections reveals a problem with WISE deblending.  About 5--10\% of WISE detections
within the SDSS imaging footprint but without a matching SDSS detection 
are blended into bright neighboring detections in WISE imaging (see Figure~\ref{fig:blending}).  In many of these cases, SDSS found
a bright extended galaxy within 10" of the WISE detection.  It appears that the WISE detection should have been matched to the SDSS galaxy,
but differences in the WISE and SDSS deblending algorithms caused a spurious offset between the WISE and SDSS positions, 
large enough so that the two detections could not be matched.  The SDSS galaxies are all contaminating $z < 0.6$ galaxies that should not
be included in our sample.  Note that we still find many of these deblending problems even when we only consider WISE detections
with NB $\leq$ 2.

In order to remove these detections from our sample, we compare 
two different measurements of $W1$ magnitude: one computed by summing all the flux within an 8.25" aperture ($W1MAG\_2$;
referred to as $A$ for clarity),
and the standard Gaussian profile-fitted magnitude measurement ($W1MPRO$; $F$ for clarity).
We expect detections with blending problems to have unusually extended flux profiles.  Since the profile-fitted magnitudes 
are largely determined by the central flux, detections with blending problems should have brighter 
8.25'' aperture magnitudes than profile-fitted magnitudes.
Indeed, the distribution of aperture
minus fitted magnitudes ($A-F$) has an asymmetrical tail of detections with an extended flux profile
(Figure~\ref{fig:blending}).
Since the distribution is nearly zero for $A-F > 0.5$, we impose a symmetrical cut on the opposite tail
and keep only those detections with 
$A-F > -0.5 $ (Figure~\ref{fig:blending}).
Visual inspection of 50 objects with $A-F < -0.5$ shows that nearly three quarters have blending issues like Figure~\ref{fig:blending}.  However, blending issues are very uncommon for $A-F > -0.5$.  We find that cutting all objects with $A-F < -0.5$ is more effective
at removing deblending problems than cutting all detections that were deblended at least once.

Our galaxy sample consists of all WISE detections meeting the above criteria that have $r - W1 > 5.5$.
Since SDSS photometry is complete to $r = 22$ and our sample only includes detections with $W1 < 16.5$,
WISE detections imaged by SDSS but lacking an SDSS match meet our color cut and are included in our sample.
Our galaxy sample consists of two components:
WISE-SDSS matches with $r - W1 > 5.5$ (``matches'') and WISE detections located within the SDSS imaging area that do not
match an SDSS detection (``non-matches''). 
In order to determine the SDSS imaging area, we use the SDSS imaging mask of \citet{ho+12},
which  is more restrictive than the SDSS imaging footprint: it removes regions
with poor seeing, low Galactic latitude, and nearby bright stars, all of which may contaminate
clustering measurements.  This yields a final area of 6966 deg$^{2}$.
We exclude all WISE detections lying outside the SDSS imaging mask of \citet{ho+12}
because we have no information about these objects' WISE-SDSS colors.

We apply additional cuts to the WISE-SDSS matches to ensure that the SDSS detections
are real objects rather than artifacts.  We remove duplicate detections by 
only including ``survey primary'' detections \citep{edr+02}.
For matches with $r < 22$, we remove
all detections with SDSS imaging flags indicating dirty photometry (e.g. saturated detections, cosmic ray strikes,
image-processing artifacts, etc.\footnote{See \url{https://www.sdss3.org/dr8/algorithms/photo\_flags\_recommend.php} for definition.}).  We do not remove detections with $r > 22$ and dirty photometry
because these detections are faint enough to be spuriously flagged as ``dirty'' even if they are real objects.
We also remove all detections with $r$-band dust extinction greater than 0.3 magnitudes.

Ignoring the WISE resolution of 6",
the astrometric precision of SDSS and WISE implies that all WISE-SDSS matches should have a separation $\lesssim$ 1". 
However, since WISE has a much larger angular resolution than SDSS,
we find many matches with separations between 1" and 3".
Visual inspection of SDSS and WISE images shows
that many matches with separations > 1" are not merely a result
of imprecise WISE astrometry.  Instead,
two SDSS detections separated by < 6" are merged
into a single WISE detection  in a substantial fraction of these matches.  If the second-closest
SDSS detection is substantially brighter than the match, it is possible that a large portion of flux from the WISE
detection should have been assigned to the second-closest SDSS detection rather than to the match.
As a result, the match may not actually meet our color cut.
We remove these problematic matches by defining the SDSS match as the brightest SDSS detection within 3" of the WISE detection.
This definition removes matches in which the second-closest SDSS detection is brighter than the match,
leading to a spuriously high $r - W1$ for that WISE detection.  By excluding these detections,
we reduce contamination in our sample.

\begin{figure*}[t]
\begin{tabular}{cccc}
\psfig{file=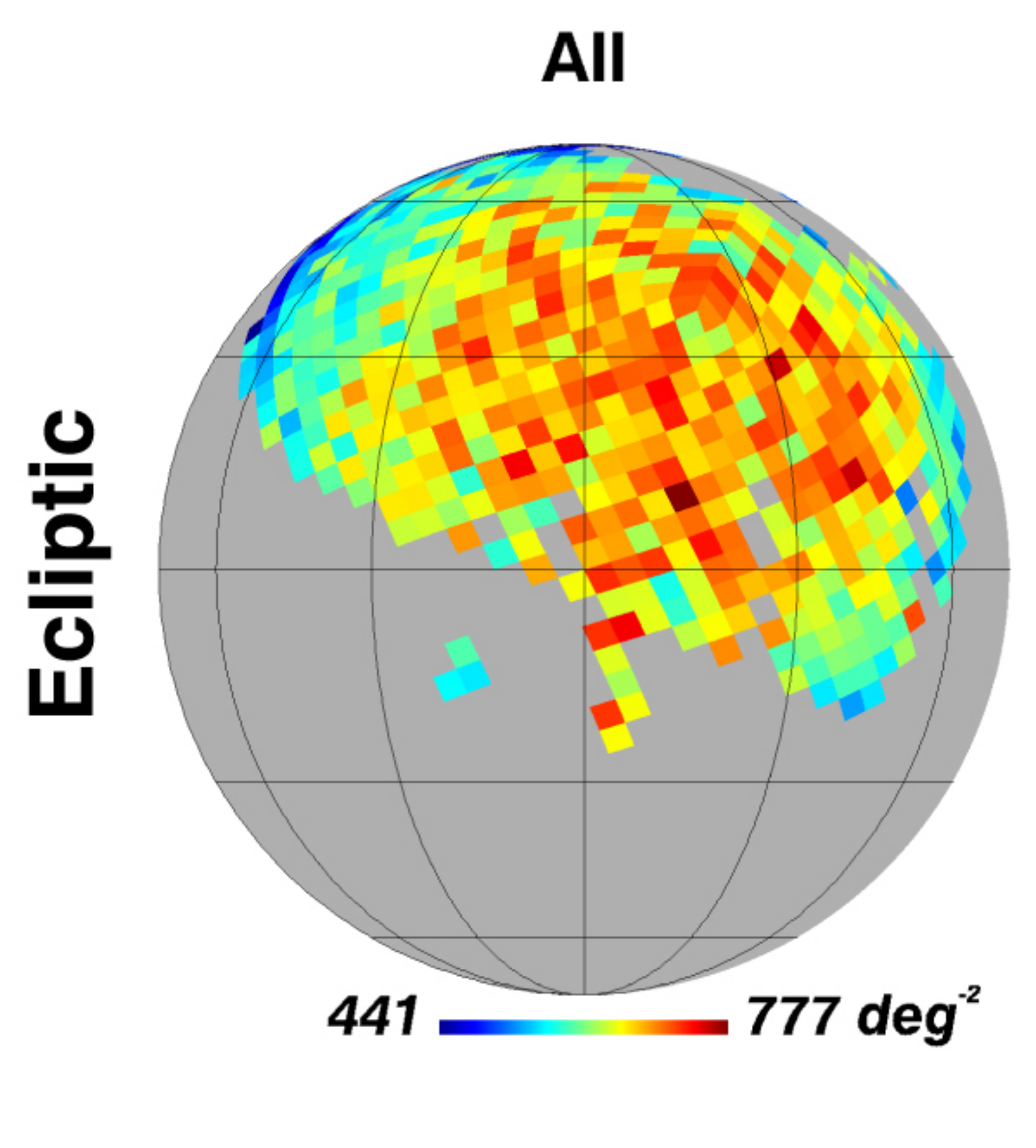,height=5.5 cm,clip=} &
\psfig{file=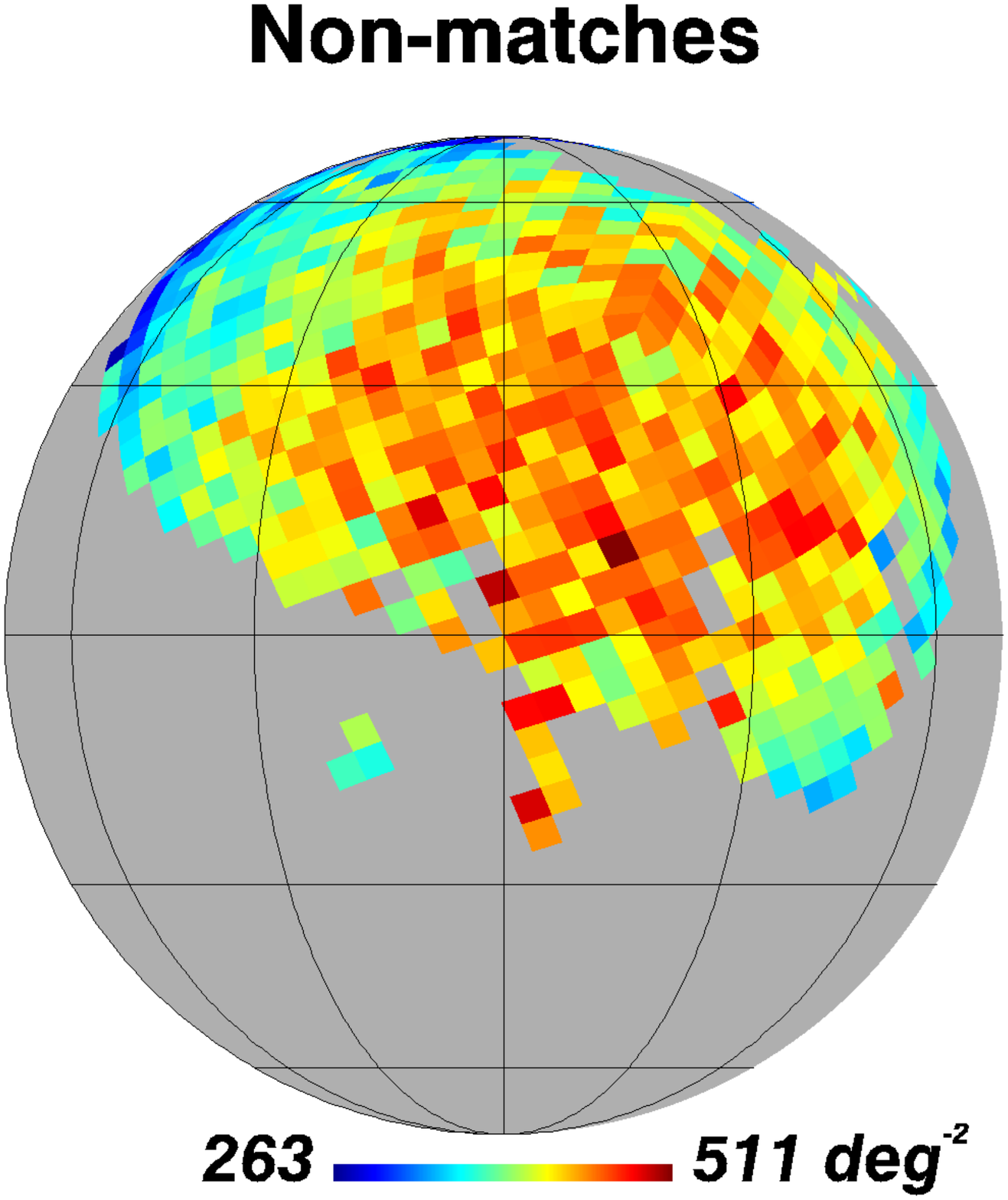,width = 4.25 cm,clip=} &
\psfig{file=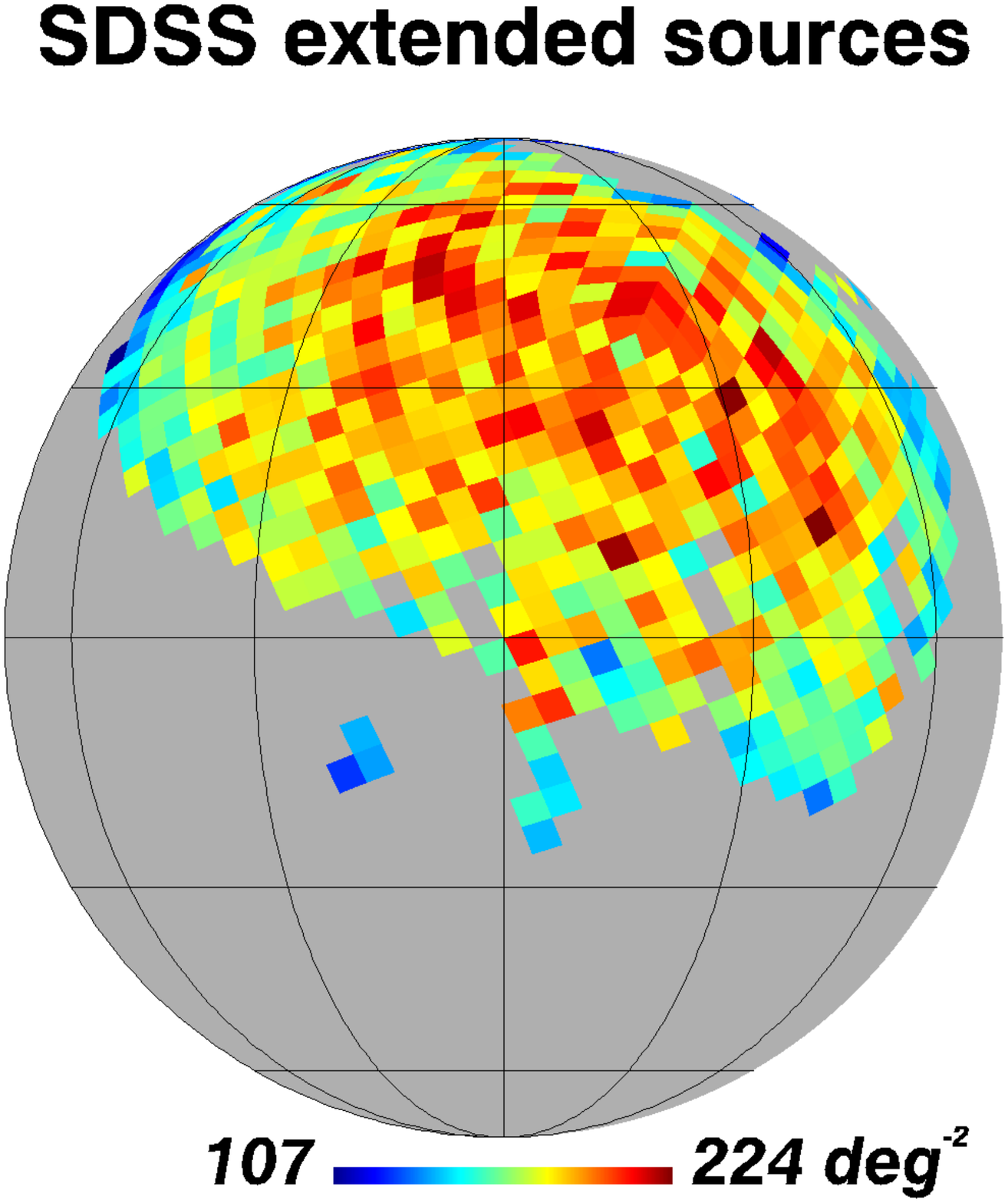,width = 4.25 cm,clip=} &
\psfig{file=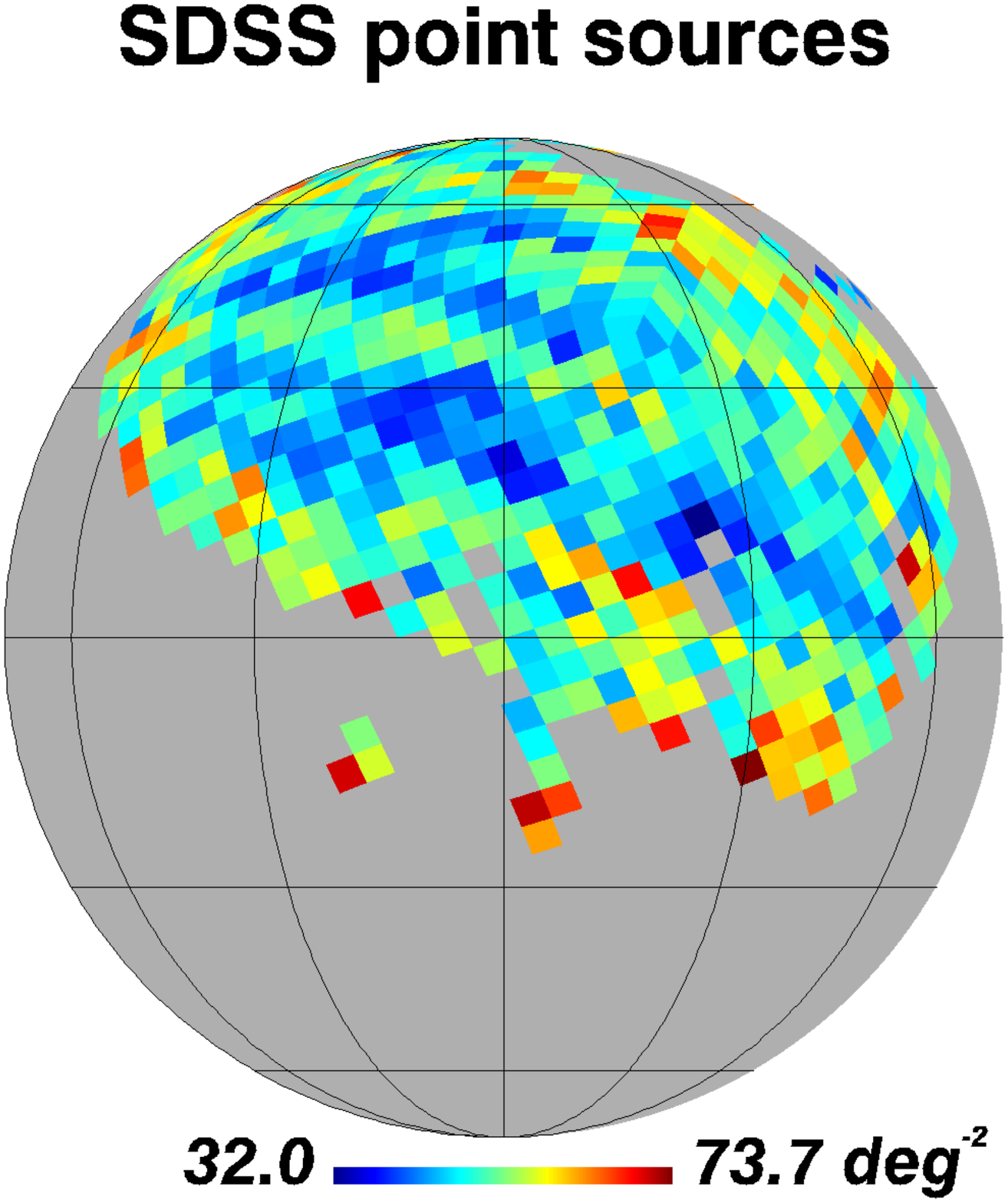,width = 4.25 cm,clip=} \\
\psfig{file=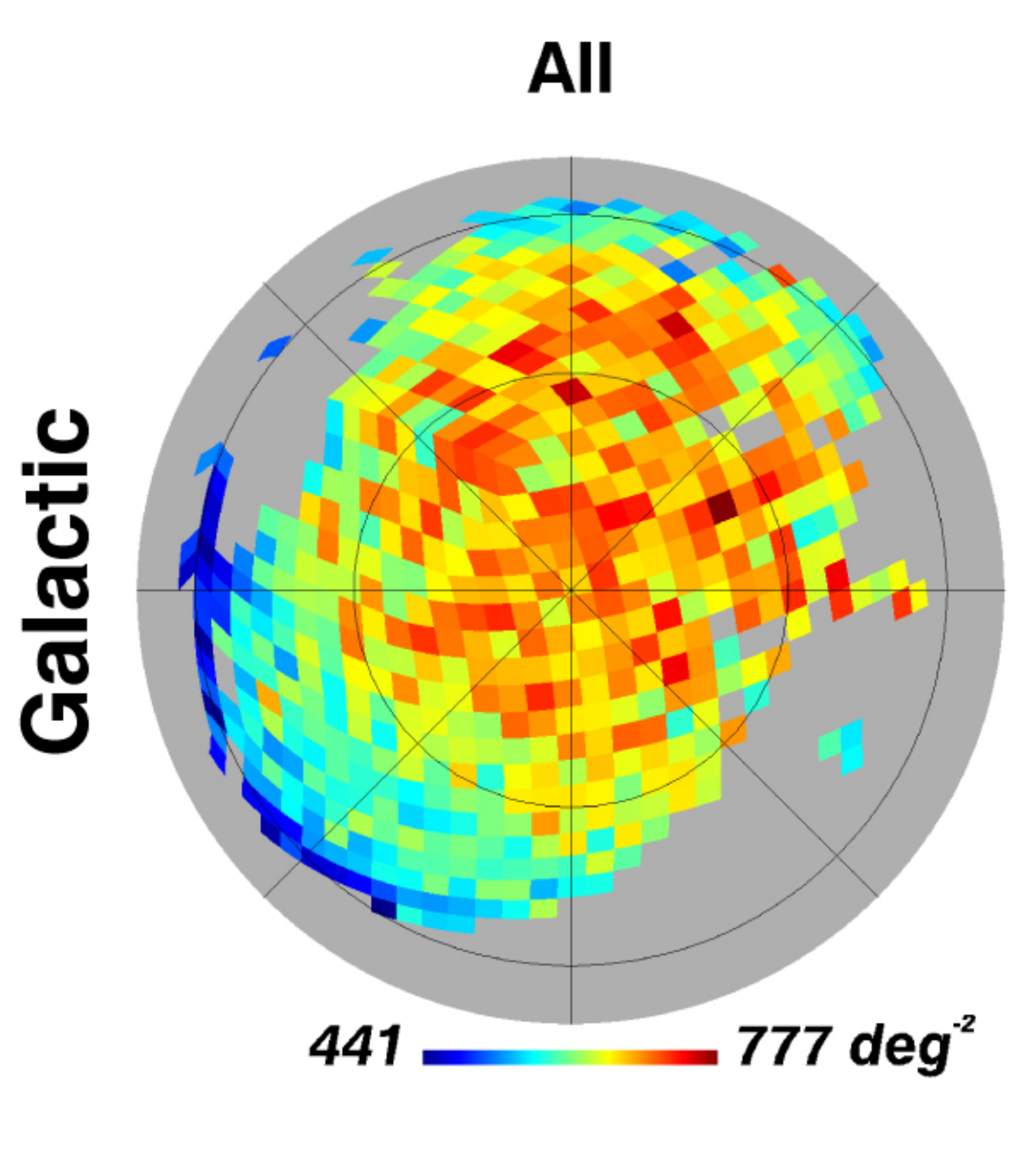,height=5.5 cm,clip=} &
\psfig{file=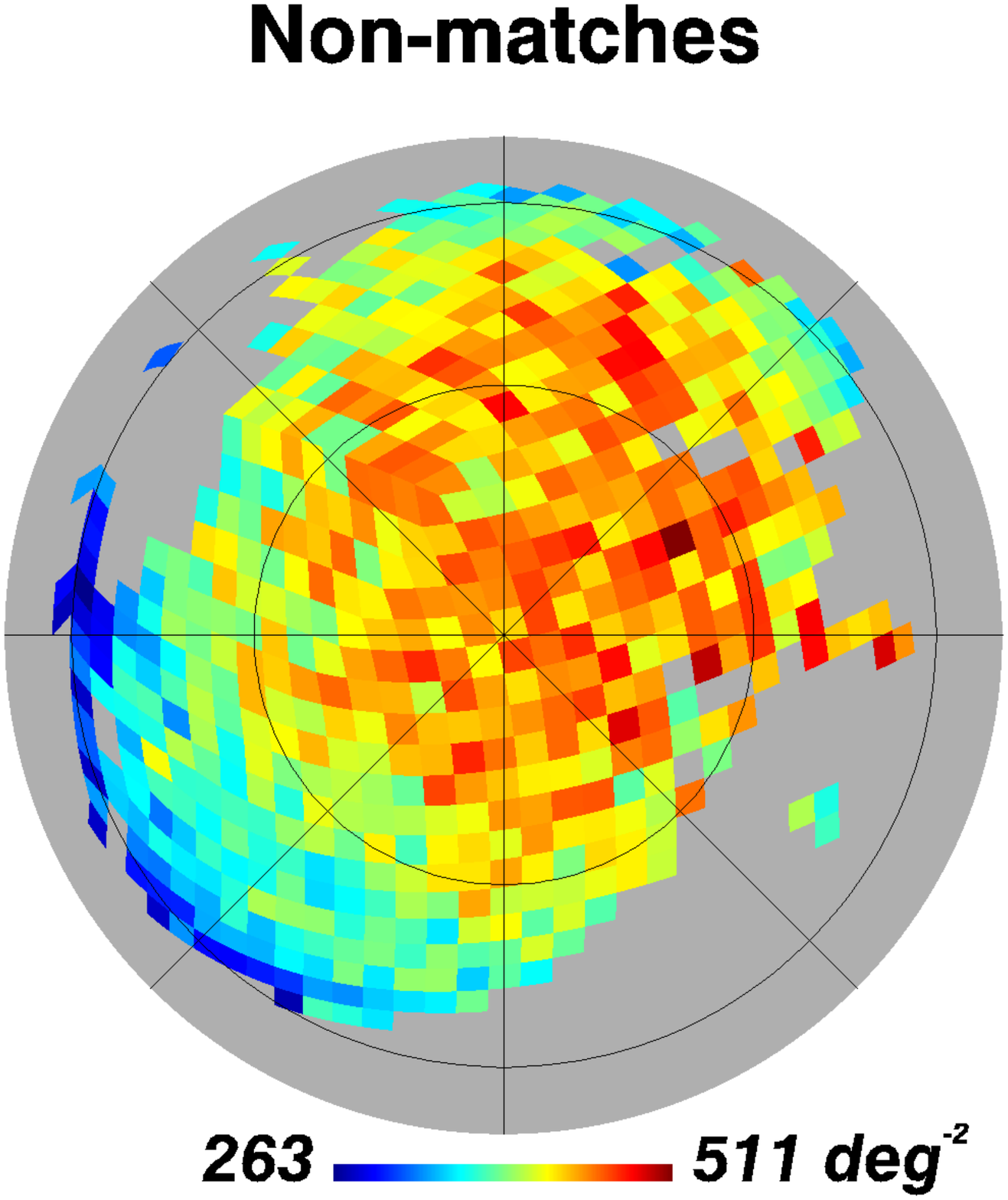,width = 4.25 cm,clip=} &
\psfig{file=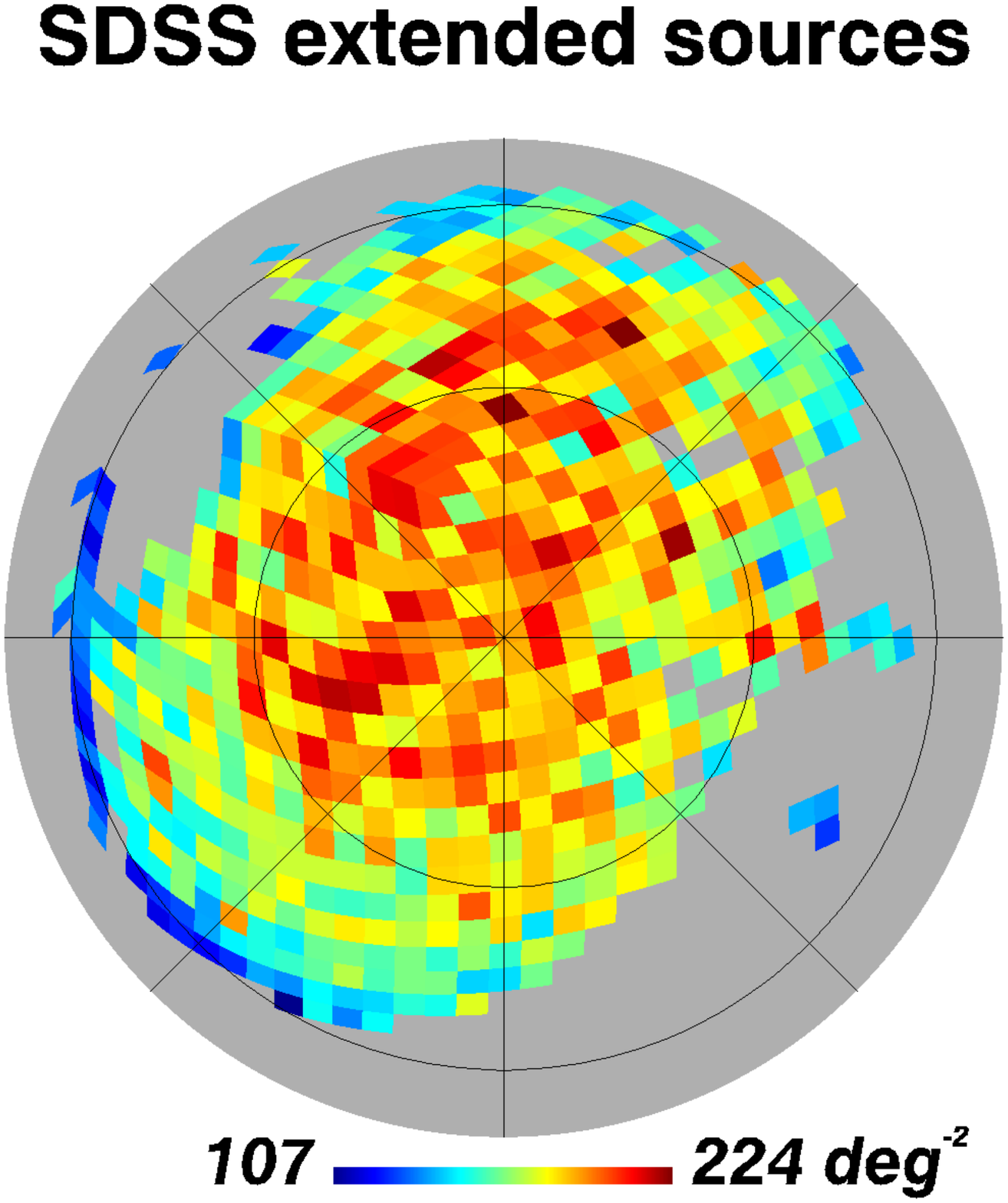,width = 4.25 cm,clip=} &
\psfig{file=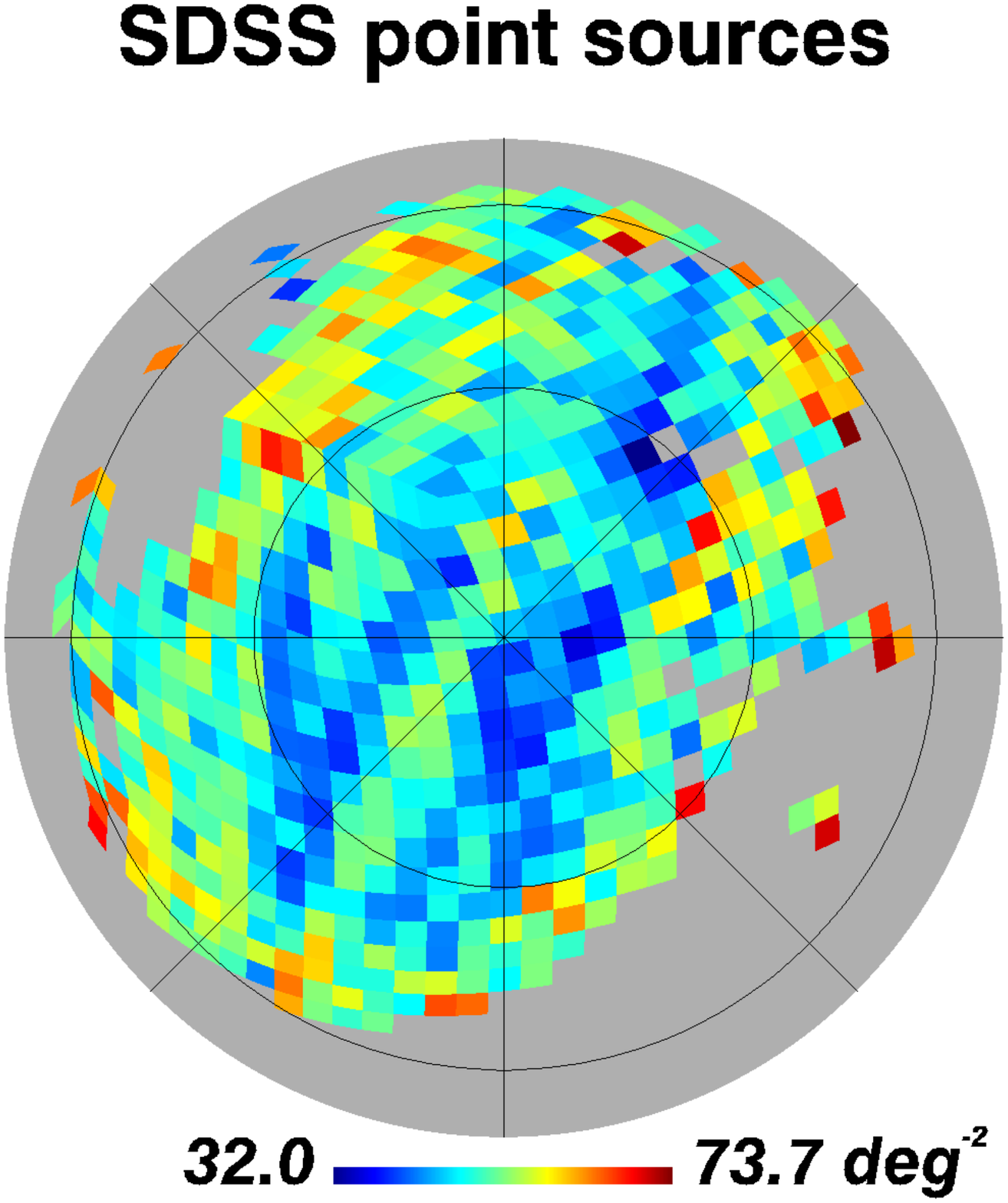,width = 4.25 cm,clip=}  
\end{tabular}
\begin{tabular}{ccc}
\psfig{file=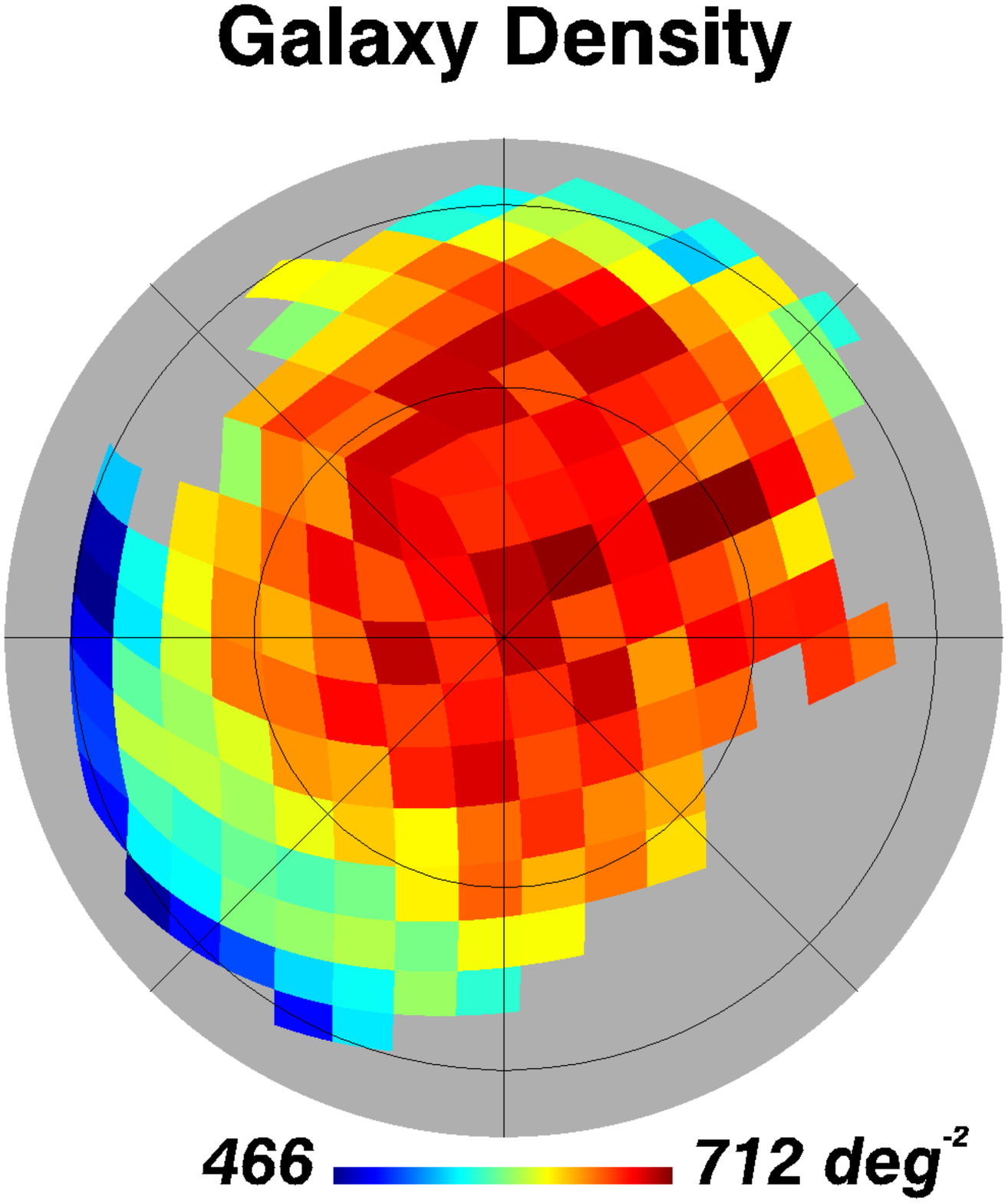,width=5.5 cm,clip=} &
\psfig{file=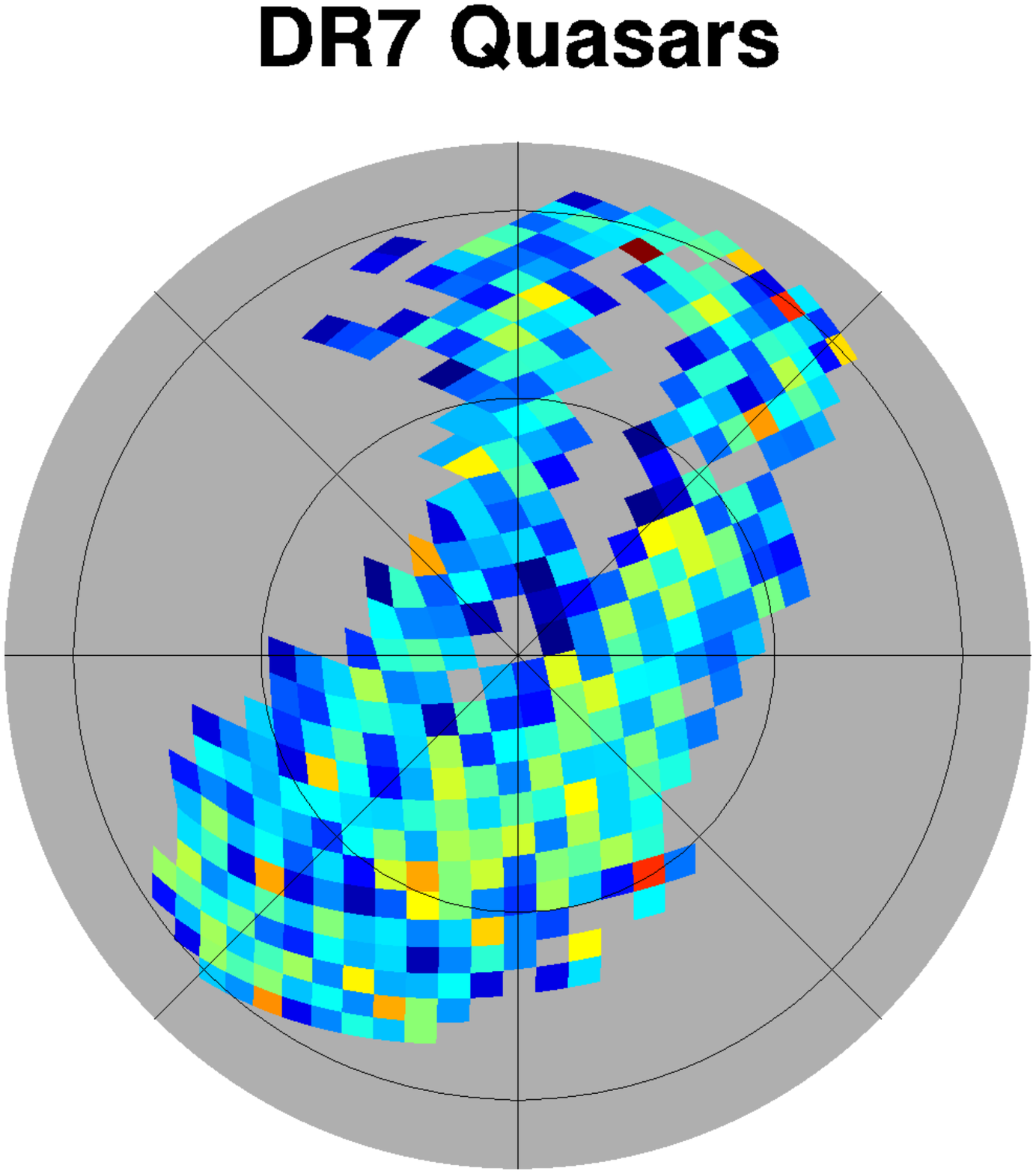,width=5.5 cm,clip=} &
\psfig{file=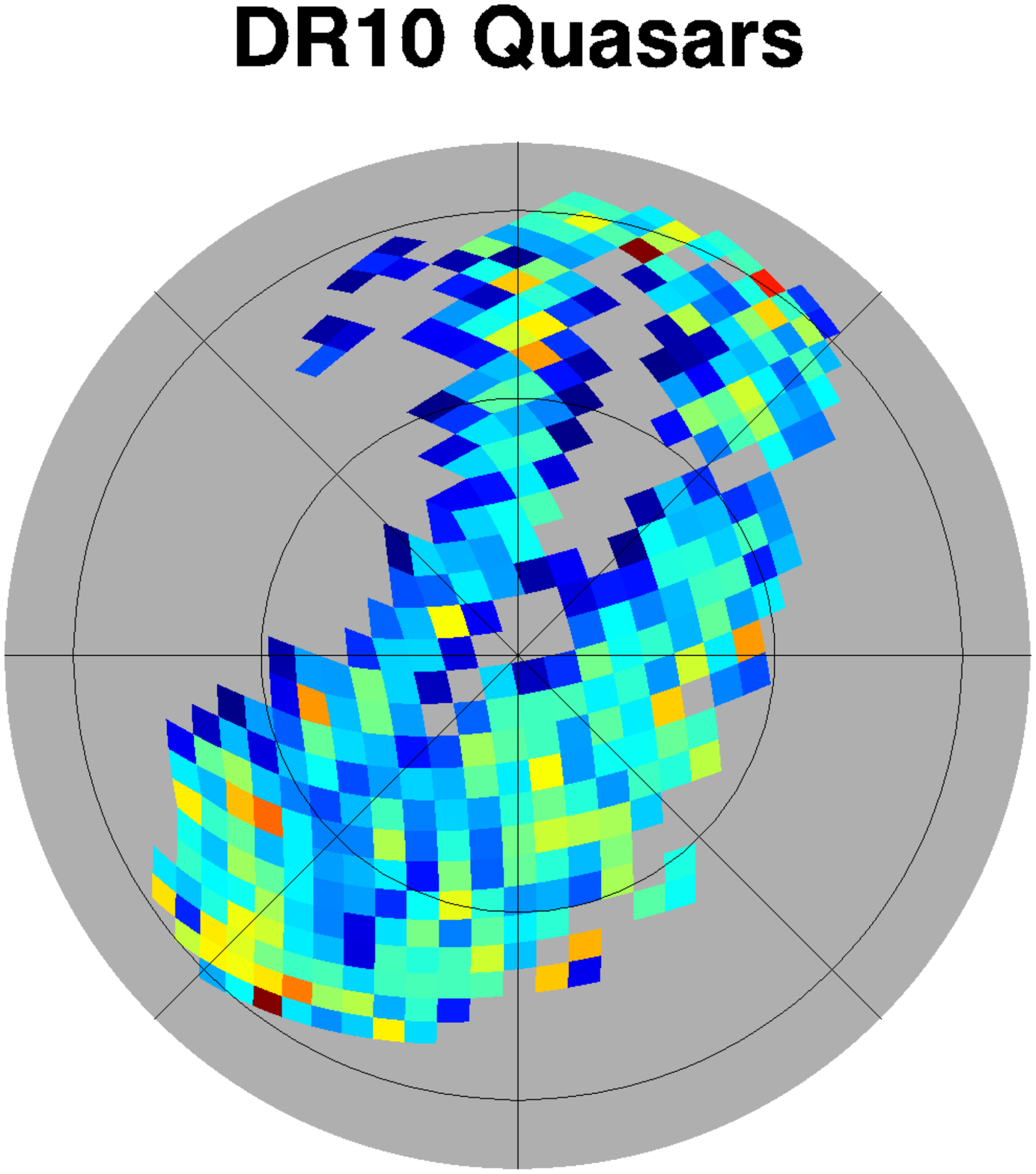,width=5.5 cm,clip=}
\end{tabular}
\caption[]{\small
\textit{Top and middle (left to right)}: Densities of the entire galaxy sample; all WISE galaxies lacking an SDSS match;
all WISE galaxies matched to an SDSS extended source; and all WISE galaxies matched to an SDSS point source.
Both plots only display HEALPix pixels with more than 20\% of their area covered by the \citet{ho+12} DR8 imaging footprint.
\textit{Top}: Ecliptic coordinates with 180$^{\circ}$ Ecliptic longitude running through the center of the plot and longitude increasing to the left.
\textit{Middle}: Galactic coordinates with
Galactic longitude increasing clockwise from the Galactic center at the bottom
of each plot.
While the variation in density with Galactic latitude is clearly apparent, we do not observe any variation in density in
regions of deeper WISE coverage (wide strip between 200$^{\circ}$ and 230$^{\circ}$ Ecliptic longitude and -90$^{\circ}$
to 90$^{\circ}$ Ecliptic latitude).
\textit{Bottom}: Comparison of galaxy and quasar sky distribution.  \textit{Bottom left}: galaxy density used in the calculation of the quasar-galaxy
angular overdensity.  Pixels with less than 20\% of their area covered by the imaging footprint are excluded.  \textit{Bottom center and right}: DR7 and DR10
quasar distribution.
Because of the low density of quasars, the fluctuations between pixels
are dominated by shot noise rather than by intrinsic variation in quasar density.  Only quasars
in the region of sky covered by both DR7 and DR10 spectroscopy are displayed.
\label{fig:sky_distribution}}  
\end{figure*}

Our final galaxy sample contains 4,168,855 objects, including 2,675,189 non-matches and 1,493,666 matches (composed of 331,327 point sources and 1,162,339 extended sources in SDSS imaging).
We estimate the severity of contamination from Galactic stars
by plotting the sky distribution of our sample in Galactic coordinates.
We also plot the sky distribution of our sample in Ecliptic coordinates
to determine whether our sample is affected by fluctuations in the WISE coverage depth
due to overlapping scans\footnote{The WISE scan pattern in is given in Figure 5 in \url{http://wise2.ipac.caltech.edu/docs/release/allsky/expsup/sec6\_2.html}.} (Figure~\ref{fig:sky_distribution}).
We observe substantial gradients on the sky related to the Galaxy
but do not observe significant gradients following the WISE scan pattern.
Both the SDSS-identified extended sources and the non-matches
have densities $\sim$80\% greater
at high Galactic latitudes.  The anti-correlation with the Galactic center results from source confusion: at low Galactic latitudes, a galaxy
is more likely to be masked by a Galactic star.
The similarity between the sky distributions of the extended sources and the non-matches suggests that the non-matches
are largely composed of $z > 0.6$ galaxies rather than Galactic stars or spurious WISE detections.

SDSS-identified point sources display the opposite pattern,
with densities nearly twice as high at low Galactic latitudes than at the poles.  This indicates that 
a substantial fraction of the point sources are very red
Galactic stars.  Nevertheless, we believe it is prudent to retain the point sources in our sample.
The SDSS star/galaxy separation is seeing dependent, with galaxy density decreasing in imaging with poor seeing,
particularly for faint detections \citep{scra+02}.  As a result, excluding the point sources from our sample would lead to seeing-dependent
spatial variations in density, which are likely correlated with seeing-dependent spatial variations in quasar density.
Given that point sources compose only 7\% of our sample and that the point sources correlate significantly with the quasars (see Figure~\ref{fig:by_class}),
we believe that is appropriate to keep the point sources in our sample to eliminate the possibility of seeing-dependent systematic error.

\subsection{Quasar selection}
\label{sec:quas_selection}

In order to obtain as large a luminosity range as possible, we measure the quasar-galaxy clustering amplitude for quasars
selected from two quasar catalogs,
the SDSS DR7 catalog with 105,783 quasars \citep{shen+11} from SDSS I/II \citep{yor+00}, and the DR10 catalog with 166,583 quasars \citep{par+13}
from BOSS \citep{eis+11,daw+13}.
In both SDSS I/II and BOSS, quasar candidates were selected from SDSS photometry using object colors.
SDSS I/II targeted objects with $15 < i < 19.1$ \citep{sch+10}, while BOSS targeted objects with either $r < 21.85$ or $g < 22$ \citep{par+13}
and $i > 17.8$ \citep{bov+11}.
Quasar candidates were arranged on spectroscopic plates \citep{blan_targ+03}, observed using a twin multi-object fiber-fed spectrograph \citep{smee+13},
and reduced and classified as a quasar, galaxy or star \citep{bolt+12}.  The wavelength range is 3800 to 9200 \AA\ for SDSS I/II and
3600 to 10400 \AA\ for BOSS \citep{smee+13}.

DR7 quasar candidates were selected using a uniform target algorithm based on their colors \citep{rich+02}.
The DR7 quasar catalog includes serendipitously imaged objects selected by other targeting algorithms,
and
we reject the half of the quasar catalog that was not selected using the uniform targeting algorithm.
Quasars in DR10 are selected using a variety of methods.  Half are selected uniformly to form the CORE sample,
using the XDQSO method \citep{bov+11}, 
and the other half are selected inhomogeneously to maximize surface density 
\bibpunct[ ]{(}{)}{,}{a}{}{;}
\citep[see][for details on the various selection algorithms]{par+13}.
\bibpunct[; ]{(}{)}{,}{a}{}{,}
We only use the CORE sample in this work.
Both the DR7 uniform quasars
and the DR10 CORE quasars are uniformly distributed across the sky.
The DR7 spectroscopic footprint is substantially larger than the DR10 footprint, because DR7 is the final data release for SDSS II while
DR10 is not the final data release for SDSS III.  We only use quasars that lie in the intersection of the DR7 and DR10 footprints.
We also only consider quasars lying within the DR8 photometric footprint of \citet{ho+12}, since our galaxy sample is restricted
to this footprint.

The final quasar sample contains 7,049 quasars with $0.65 < z < 0.9$, 4,206 from DR10 and 2,843 from DR7.
Figure~\ref{fig:z_bh_luminosity} shows the redshift, luminosity and virial mass
distributions for the DR7 and DR10 quasars.  To measure
the luminosity dependence of the clustering amplitude, we split the DR7 quasars into three groups by luminosity and the DR10 quasars into four groups by luminosity.
Similarly, we split the DR7 quasars into four groups by virial mass and the DR10 quasars into four groups by virial mass.

Redshifts in both the DR7 and DR10 catalogs are accurate to $\Delta z < 0.01$ \citep{sch+10,par+13}.  In this paper,
we measure luminosity using the absolute $i$-band magnitude K-corrected to $z = 2$ for both DR7 and DR10 \citep{rich+06}.
These magnitudes
measure the quasar luminosity in a bandpass centered at 2500 \AA\ in the rest frame.  We convert absolute magnitude to 2500 \AA\ luminosity
in erg s$^{-1}$ using Equation 4 from \citet{rich+06} and then to solar luminosity using $L_{\odot} = 3.827 \times 10^{33}$ erg s$^{-1}$.
Reported $i$-band magnitude errors for quasars in our sample are $\leq 0.03$ for DR7 and $\leq 0.1$ for DR10.  
Note that the
DR7 quasars are substantially more luminous than the DR10 quasars (Figure~\ref{fig:z_bh_luminosity}), since BOSS targeted
fainter, higher-redshift quasars compared to SDSS I/II.

We use the single-epoch virial mass estimates from \citet{shen+11}
for DR7 quasars, and estimates computed using similar methodology for the DR10 quasars (Y. Shen, private communication)
\footnote{The line fitting methodology for DR10 is described in \citet{shen_liu+12} and at \url{http://users.obs.carnegiescience.edu/yshen/BH_mass/dr9.htm}.
The practical impact of the different fitting procedures for DR7 and DR10 is negligible (Y. Shen, private communication).}.
The virial mass estimates assume that the quasar's broad-line region (BLR) is virialized:
\begin{equation}
M_{\mathrm{enc}} = M_{\mathrm{BH}} = \frac{V_{\mathrm{BLR}}^2 R_{\mathrm{BLR}}}{G}
\label{eqn:virial_mass}
\end{equation}
The BLR velocity is inferred from the width of a particular broad line.
The radius is determined by measuring the continuum luminosity:
reverberation mapping of $z < 0.3$ AGN found a tight relationship between continuum luminosity and BLR radius \citep{kas+00}.
This relationship arises because the size of the BLR is regulated by the amount of ionizing radiation emitted by the quasar,
which is proportional to the optical continuum luminosity.
Since the reverberation mapping samples used to calibrate the single-epoch virial mass estimates use H$\alpha$ and H$\beta$ line widths,
virial mass estimators based on H$\alpha$ and H$\beta$ are most reliable \citep{shen_mass+13}.  However, H$\beta$ is redshifted beyond the edge
of the DR7 spectrograph (9200 \AA) at $z > 0.85$, so we instead use MgII based mass estimates.
MgII line widths correlate well with H$\beta$ line widths \citep{shen_mass+13}, but the 
MgII masses may possess substantial systematic errors due to the lack of MgII-based reverberation mapping masses.  However, since
we are interested in the slope rather than the normalization of the black hole mass-clustering strength relationship, 
we are not concerned with systematic errors resulting in a uniform offset in black hole
mass.

We are primarily concerned with two kinds of error in the virial mass estimates: increased scatter due to the large uncertainties in the virial
mass estimates and a luminosity-dependent bias arising from the flux limit of the sample.
First,
the spread in virial mass is wider than the spread in true mass
because of the large uncertainties in individual virial mass estimates.  \citet{shen_mass+13} estimated that the scatter in virial mass
at fixed true mass is up to $\approx$ 0.5 decades, arising from both measurement errors in line width and continuum luminosity, and uncertainties in the virial
mass calibrations.  We reduce the measurement uncertainties slightly by excluding quasars with poor continuum or MgII emission line fits ($\chi^2$/d.o.f > 2),
but this does not reduce the calibration uncertainties, which are the dominant source of scatter in virial mass at fixed true mass.  Because of the large uncertainty
in virial mass,
the dynamic range of true mass in our sample is less than the measured range in virial mass, and the mean true mass for quasars in a given virial mass bin is
less extreme than the mean virial mass.

\citet{shen_mass+13} also discussed a luminosity-dependent bias arising from uncertainties
in line width and luminosity.  Measurement errors, scatter in the radius-luminosity relationship,
non-virial motion,
and a time lag between changes in luminosity and radius may lead to uncorrelated errors in luminosity and line width.
If uncorrelated errors are present, virial masses in flux-limited samples will be biased high relative to
true masses, since
the increase in average luminosity caused by the flux limit will not be entirely cancelled by a decrease in line width.
The magnitude of the bias will be greatest for low luminosity subsamples that lose a substantial number of quasars to the flux limit.
\citet{shen_mass+13} and \citet{sk+12} found evidence for luminosity-dependent bias in MgII masses.
Since less massive quasars are also less luminous,
the luminosity-dependent bias will be stronger for low mass quasars than for high mass quasars.  However, since each bin in virial mass covers
a large range in true mass, the magnitude of the luminosity-dependent bias is quite similar for all virial mass bins.
Because our quasar samples are flux limited, the luminosity-dependent bias causes the virial masses to systematically overestimate the true masses,
but this uniform offset in black hole mass does not affect our measurement of the slope of the black hole mass-clustering strength relationship.

\subsection{Systematic effects}
\label{sec:syst}

The difference in redshift distribution between different quasar groups leads to systematic differences in quasar-galaxy clustering strength.
The quasar luminosity function evolves significantly with redshift \citep{rich+06}, so high luminosity quasars have a higher mean redshift
than low luminosity quasars.  
Angular clustering strength varies with redshift due to both redshift evolution of the quasar autocorrelation function \citep{cro+05}
and the redshift distribution of our galaxy sample.  
Therefore, any difference in clustering between quasar groups with different luminosities may in fact arise from the difference in redshifts.

\begin{figure*}[t]
\begin{tabular}{ccc}
\hspace{-15pt}
\psfig{file=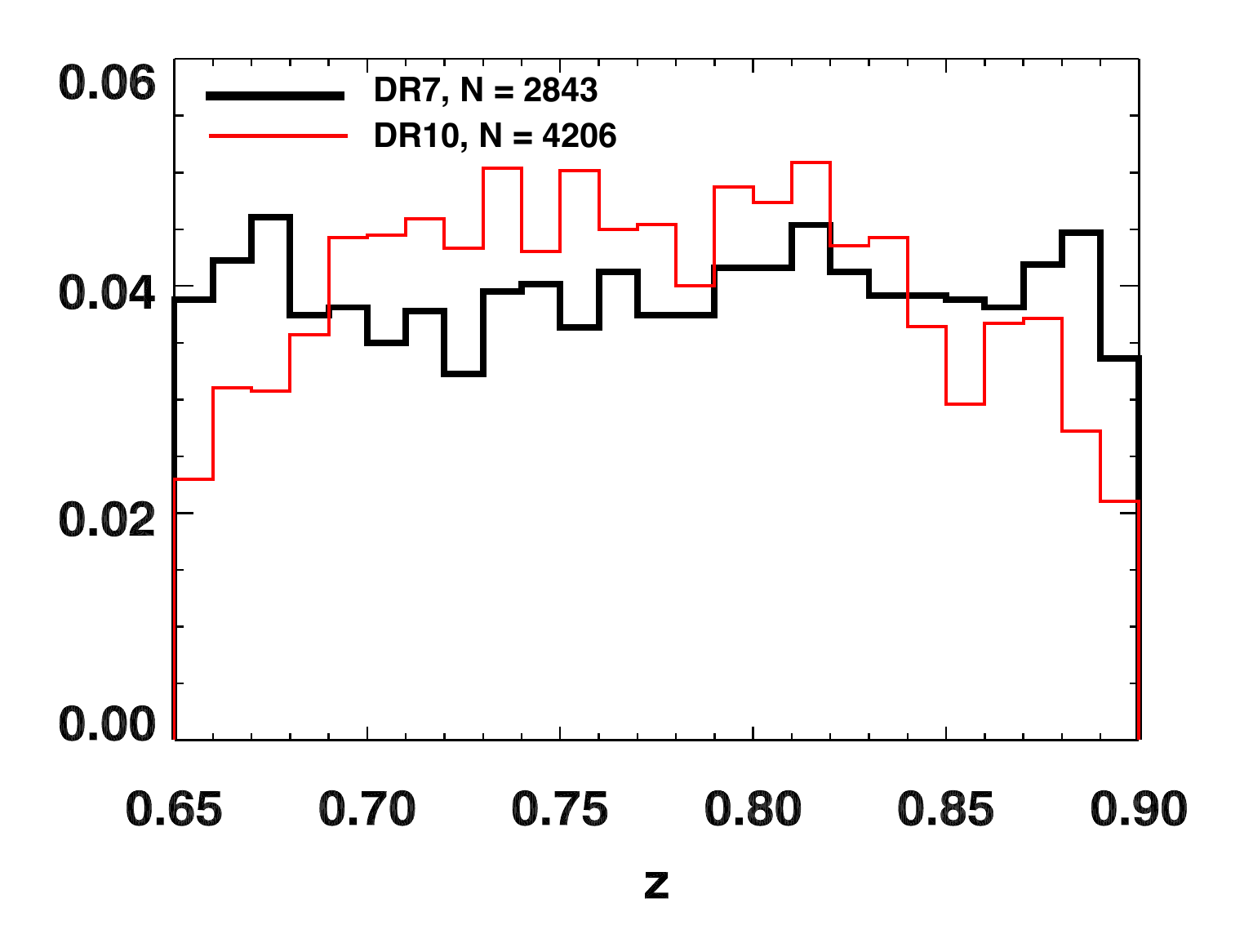,height=4.8cm, clip=} &
\hspace{-20pt}
\psfig{file=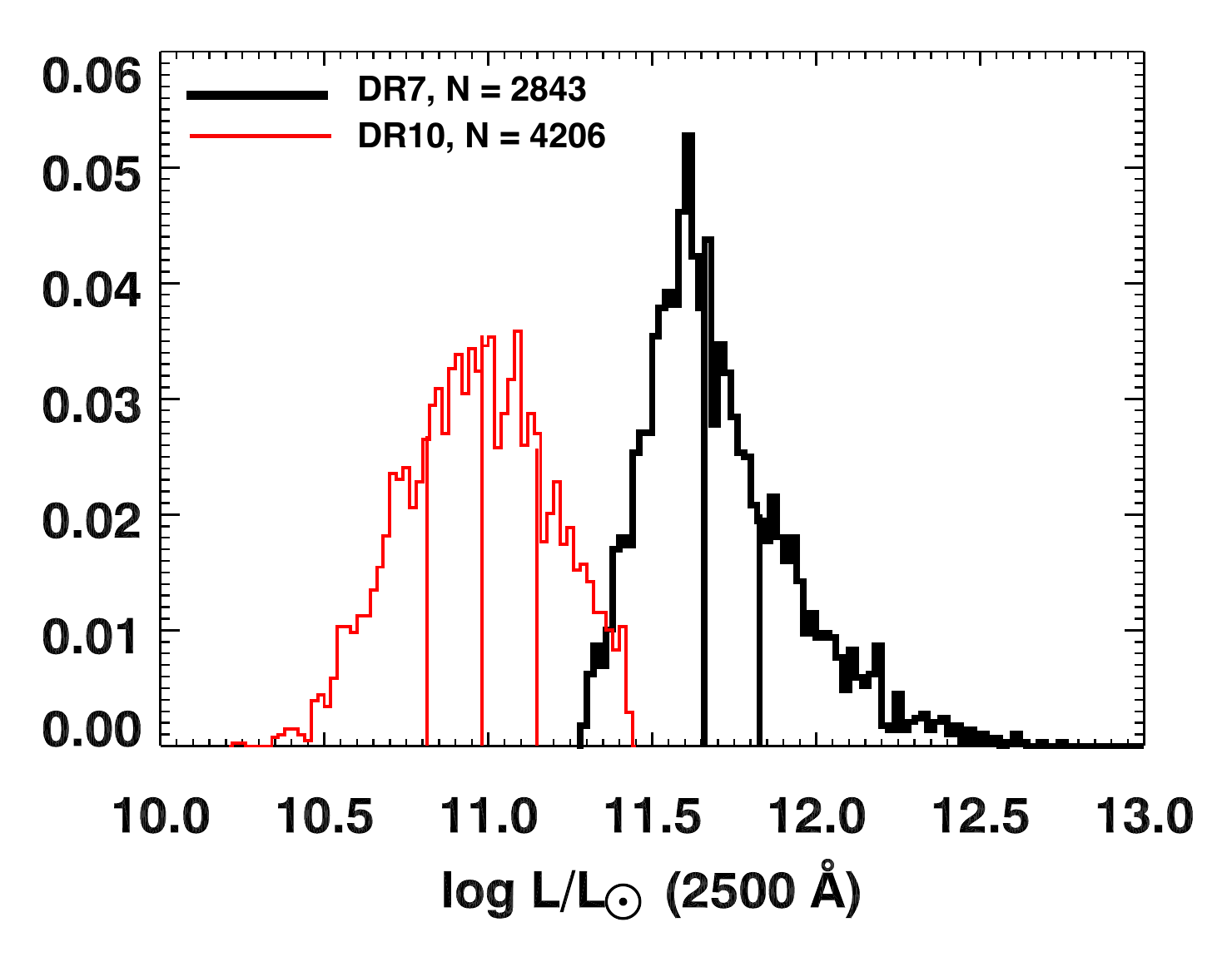,height =4.8cm, clip=} &
\hspace{-20pt}
\psfig{file=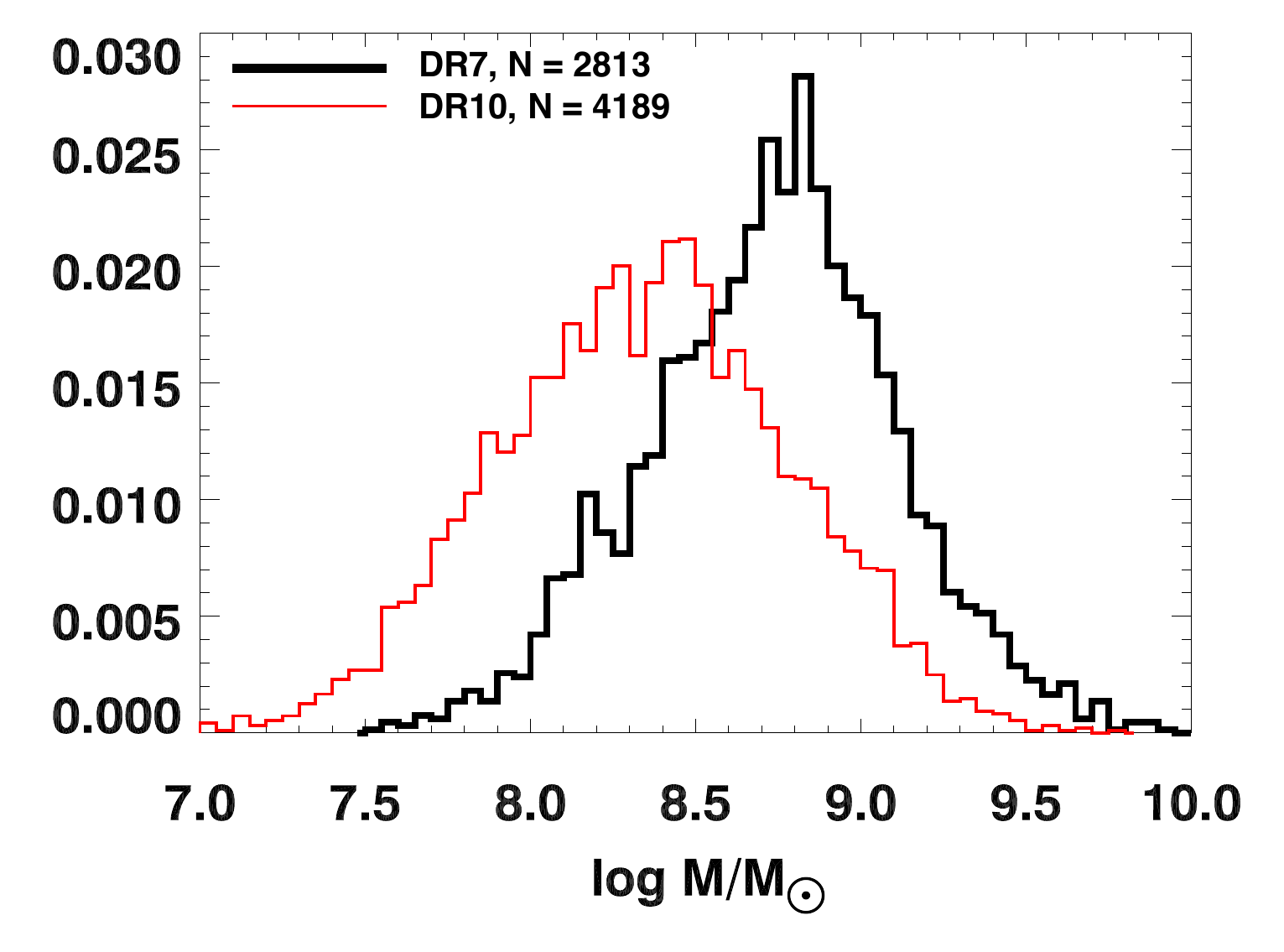, height =4.8cm, clip=}
\end{tabular}

\caption[]{\small \textit{Left}: redshift distribution for DR7 and DR10 quasars.
\textit{Center}: Distributions of $M_{i}(z=2)$ for DR7 and DR10 quasars ($\approx 2500$ \AA\ restframe wavelength) in solar luminosities.  The vertical lines divide the sample into 4 DR10 groups
and 3 DR7 groups.  The right-most cut
in the DR10 distribution at $11.43 \log{L/L_{\odot}}$ minimizes the overlap between the most luminous DR10 sample and the least luminous DR7 sample.
\textit{Right}: Virial black hole masses for DR7 and DR10 quasars, computed by \citet{shen+11} from continuum luminosity and emission line widths.  The DR7 and DR10 quasars
are each divided into 4 groups; the divisions are not shown because several of the bins overlap.
\label{fig:z_bh_luminosity}}  
\end{figure*}

To isolate the luminosity dependence of clustering,
we assign a redshift-dependent weight to each quasar to force the weighted redshift distributions of each subsample
to match the DR10 redshift distribution.  For each subsample, we place the quasars into bins of width $\Delta z = 0.01$
and compute the weights by dividing the DR10 redshift distribution by the subsample's redshift distribution.
To ensure that no quasars are weighted by more than 3 or less than 1/3, we combine the two lowest-luminosity DR7 groups (see Figure~\ref{fig:z_bh_luminosity}).

Because the density of both the WISE galaxies and the stellar contaminants in the WISE sample vary
across the sky, differences in large-scale sky distribution may lead to differences in clustering amplitude.
Using only quasars located within the DR7/DR10 overlap region largely alleviates this problem by forcing the sky distribution of DR7 and DR10 quasars to match
(see Figure~\ref{fig:sky_distribution}).
While the resulting sky distributions are not identical, the residual variation in both WISE galaxy density
and contaminating star density is quite small.  We eliminate the effects of residual variation in WISE galaxy density by measuring
the galaxy density separately in each HEALPix pixel (see \citet{gor+05} for details about HEALPix).
The stellar contamination, as measured by the average fraction of SDSS point sources in the HEALPix pixel surrounding each quasar,
is just 1.2\% greater for DR10 quasars than for DR7 quasars.  The impact of varying stellar contamination is therefore considerably lower than
the magnitude of the measured overdensity $w$.

Small scale variations in galaxy density may also affect our clustering measurements.  While Figure~\ref{fig:sky_distribution} cannot
display density variations on scales less than a few degrees,
we expect to observe variations in WISE density on arcminute scales due to both source suppression near bright sources
and image artifacts within the halos of very bright sources\footnote{
See \url{http://wise2.ipac.caltech.edu/docs/release/allsky/expsup/sec6\_2.html}.}.  The vast majority of these bright sources are uncorrelated
with quasar positions, so we do not apply a mask to remove the area around bright WISE sources.  
Unlike the large-scale variations in quasar and galaxy densities, small-scale variations in galaxy density should have the same
effect on both DR7 and DR10 groups, and on different
groups in luminosity and virial mass.
However, bright quasars in WISE imaging may lead to both an overdensity of nearby image artifacts and an underdensity of nearby faint sources
due to the increased background.
While we failed to find image artifacts around the 50 brightest quasars in $W1$, these quasars, with $W1 \approx 12$,
suppress
the density of $W1 = 16.5$ sources at separations < 18"\footnote{See Figure 25 at \url{http://wise2.ipac.caltech.edu/docs/release/allsky/expsup/sec6\_2.html}.}.  
As a result, we restrict our measurement of the quasar-galaxy cross-correlation function to angular scales > 18", corresponding to $\approx 0.2$ $h^{-1}$ Mpc.
At these scales, we expect no significant suppression of galaxy density due to increased background from nearby quasars.
Similar source suppression in SDSS is only observed at separations less than 15" for galaxies with similar brightness as our galaxy sample
about stars with similar brightness as our quasar sample \citep{ross+11}.

\section{Measuring the angular overdensity}
\label{sec:angular_overdensity}

\subsection{Methods}
\label{sec:methods}

In this paper, we measure the angular overdensity $w(r_p)$ rather than the three-dimensional correlation
function $\xi(r)$ or the projected cross-correlation function $w_p(r_p)$.
The three-dimensional quasar-galaxy cross-correlation function is well-fit by a power law \citep[e.g.][]{yg+87}:
\begin{equation}
\xi(r) = \left(\frac{r_0}{r} \right)^{\gamma}
\label{eqn:ccf}
\end{equation}
where $r_0$ is the correlation length, which is typically used to characterize clustering strength.
The projected cross-correlation function $w_p(r_p)$ is the integral of the three-dimensional cross-correlation function along the line of sight:
\begin{equation}
w_p(r_p) = \int_{-\infty}^{\infty} \! \xi(r_p,\pi) \, \mathrm{d}\pi = r_{0}^{\gamma} \frac{\Gamma(1/2) \Gamma[(\gamma - 1)/2]}{\Gamma(\gamma/2)} r_{p}^{1-\gamma}
\label{eqn:proj_ccf}
\end{equation}
where $\Gamma$ is the gamma function and $\pi$ is the line-of-sight distance.
Following \citet{zh+13}, we can relate
the projected cross-correlation function $w_p(r_p)$ to the angular overdensity $w(r_p)$:
\begin{equation}
w_p (r_p) = \left \langle \frac{ n}{\rho_{0}} \right \rangle w(r_p)
\label{eqn:wtoomega}
\end{equation}
where
$n$ is the WISE density near each quasar (Figure~\ref{fig:sky_distribution}) and $\rho_{0}$ is the density of galaxies 
at each quasar's redshift that meet our color and $W1$ cuts.  

Estimating $r_0$ and $b_Q$ requires finding the projected cross-correlation function, which requires an estimate of $\rho_0$.
One can compute $\rho_0$ by integrating the galaxy luminosity function at each redshift \citep[e.g.][]{zh+13,komi+13}.
However, this method is insufficiently precise to measure the host halo mass to a reasonable accuracy: since bias
is a shallow function of host halo mass at $z \approx 0.8$, small errors in the luminosity function (and thus $\rho_0$, $r_0$, and the bias)
lead to relatively large errors in the host halo mass.  For instance, small differences in photometric calibration between SDSS
and the instrument used to find the luminosity function will lead to substantial errors in the luminosity function,
causing large uncertainties in the host halo mass.  Moreover, to find $\rho_0$ we would need to apply our color cut
to the galaxy luminosity functions, requiring knowledge of galaxy SEDs at $z \approx 0.8$.  Ultimately, finding $r_0$ and $b_Q$
is not necessary to measure the change in clustering with luminosity and virial mass, since such changes will be just as apparent in
the angular overdensity $w(r_p)$.
Given the difficulties associated with computing $r_0$ and $b_Q$ from our data, we choose to only measure the clustering
dependence of $w(r_p)$ in this paper.

We measure the angular quasar-galaxy overdensity
by counting the number of excess galaxies at angular separation $\theta$ from each quasar.
We count the number of galaxies $N$ in an annulus of width $\Delta \theta$
and divide by the expected density of galaxies:
\begin{equation}
w(\theta) = \left(\sum_{i} w_{i} \frac{N_{i}}{n_{i} A_{i}}\right) \bigg/ \left(\sum_{i} w_{i}\right) - 1
\label{eqn:w1}
\end{equation}
where $n$ is the average galaxy density, $A$ is the area inside the SDSS imaging footprint,
the index $i$ ranges over every quasar, and the weights $w$ are computed from the redshift distribution as described in Section~\ref{sec:quas_selection}.
Because the galaxy density varies 
substantially across
the sky, $n$ is estimated by using the galaxy density in each HEALPix pixel (Figure~\ref{fig:sky_distribution}).
We exclude 18 quasars lying in pixels with less than 20\% coverage of the imaging footprint; for these pixels, Poisson variations in the density are > 1\%,
comparable in magnitude to the Galactic gradients shown in Figure~\ref{fig:sky_distribution}, 
and thus we consider the density estimates for these quasars to be unreliable.

We determine $A$, the area within the imaging mask, using a Monte Carlo method.  
We create a catalog of random points lying outside the imaging footprint with density $n_{\textrm{random}}$ 2909 deg$^{-2}$,
much larger than the maximum 
$n$, 712 deg$^{-2}$
(Figure~\ref{fig:sky_distribution}).  We use this catalog to find the area within the imaging footprint, $A$, by subtracting the
area outside the imaging footprint from the area of the full annulus, $A_{\textrm{full}}$:
\begin{equation}
A = A_{\textrm{full}} - \frac{N_{\textrm{random}}}{n_{\textrm{random}}}
\label{eqn:area}
\end{equation}
where the area outside the imaging footprint is determined by dividing the number of random points in the annulus, $N_{\textrm{random}}$,
by the density of random points $n_{\textrm{random}}$.
Next we transform from  $w(\theta)$ to $w(r_p)$, where $r_p$, measured in comoving
$h^{-1}$ Mpc, is the distance between the quasar
and the galaxy assuming that the galaxy lies at the same redshift as the quasar.
We compute the area of the full annulus, A$_{\textrm{full}}$, using $r_p$ rather than $\theta$:
\begin{equation}
A_{\textrm{full}} = \pi \frac{r_{2}^2 - r_{1}^2}{((1+z)D_{A})^{2}}
\label{eqn:afull}
\end{equation}
where $r_{1} = r_p - \Delta r_p$, $r_{2} = r_p + \Delta r_p$, and $D_{A}(z)$ is the angular-diameter distance to a quasar
at redshift $z$.  

We measure $w(r_p)$ in the following 5 bins (in units of $h^{-1}$ Mpc): 0.2--0.4, 0.4--0.8, 0.8--1.6, 1.6--3.2, and 3.2--6.4.  
We cannot measure $w(r_p)$ at separations less than 0.2  $h^{-1}$
Mpc (angular separation 18") because at smaller separations the density of $W1$ = 16.5 galaxies is suppressed by the wings of the central quasar's flux profile.
At angular scales larger than 6.4 $h^{-1}$ Mpc, we find that our measurement of $w(r_p)$ is contaminated by systematic errors (see Section~\ref{sec:systematics}).

We use bootstrap resampling to estimate our error bars.  We resample by HEALPix pixel:
from the 154 pixels with at least one quasar, we randomly select 154 pixels with replacement and measure $w(r_p)$ for all quasars in the selected pixels.
We use 50,000 resamples
to calculate error bars and the covariance matrix.  Equation~\ref{eqn:cov_matrix} gives
the reduced covariance matrix for the measurement of $w$ across the entire sample, $R_{ij} = C_{ij}/ \sqrt{C_{ii}C_{jj}}$, where $C$ is the covariance matrix:

\begin{equation}
R = \scalebox{0.9}{$\begin{pmatrix}
1.000 & 0.234 & 0.230 & 0.152 & -0.005  \\
0.234 & 1.000 & 0.336 & 0.226 & 0.025  \\
0.230 & 0.336 & 1.000 & 0.306 & 0.220  \\
0.152 & 0.226 & 0.306 & 1.000 & 0.595  \\
-0.005 & 0.025 & 0.220 & 0.595 & 1.000
\end{pmatrix}$}
\label{eqn:cov_matrix}
\end{equation}

Since $w_p(r_p)$ is related by $w(r_p)$ by a constant of proportionality  (Equation~\ref{eqn:wtoomega}),
we fit a power law to $w(r_p)$:
\begin{equation}
w(r_p) = \beta r_p^{-\delta}
\label{eqn:w}
\end{equation}
where $\beta \propto r_{0}^{\gamma}$ and $\delta = \gamma - 1$.
We use $\chi^{2}$ minimization to find the best fit values of $\beta$ and $\delta$ for our measurement of $w(r_p)$
using the entire quasar and galaxy samples.  Since the off-diagonal terms of the covariance matrix are nonzero (Equation~\ref{eqn:cov_matrix}), we 
compute $\chi^{2}$ using the full covariance matrix.
Using bootstrapping, we confirm that the sampling distribution of $w(r)$ in each bin is very well approximated by a Gaussian, indicating
that $\chi^2$ minimization is an appropriate curvefitting technique.
We will 
use the clustering amplitude $\beta$ to characterize the strength of quasar-galaxy clustering for each of our subsamples.

\begin{figure*}[t]
\begin{tabular}{cc}
\hspace{-20pt}
\psfig{file=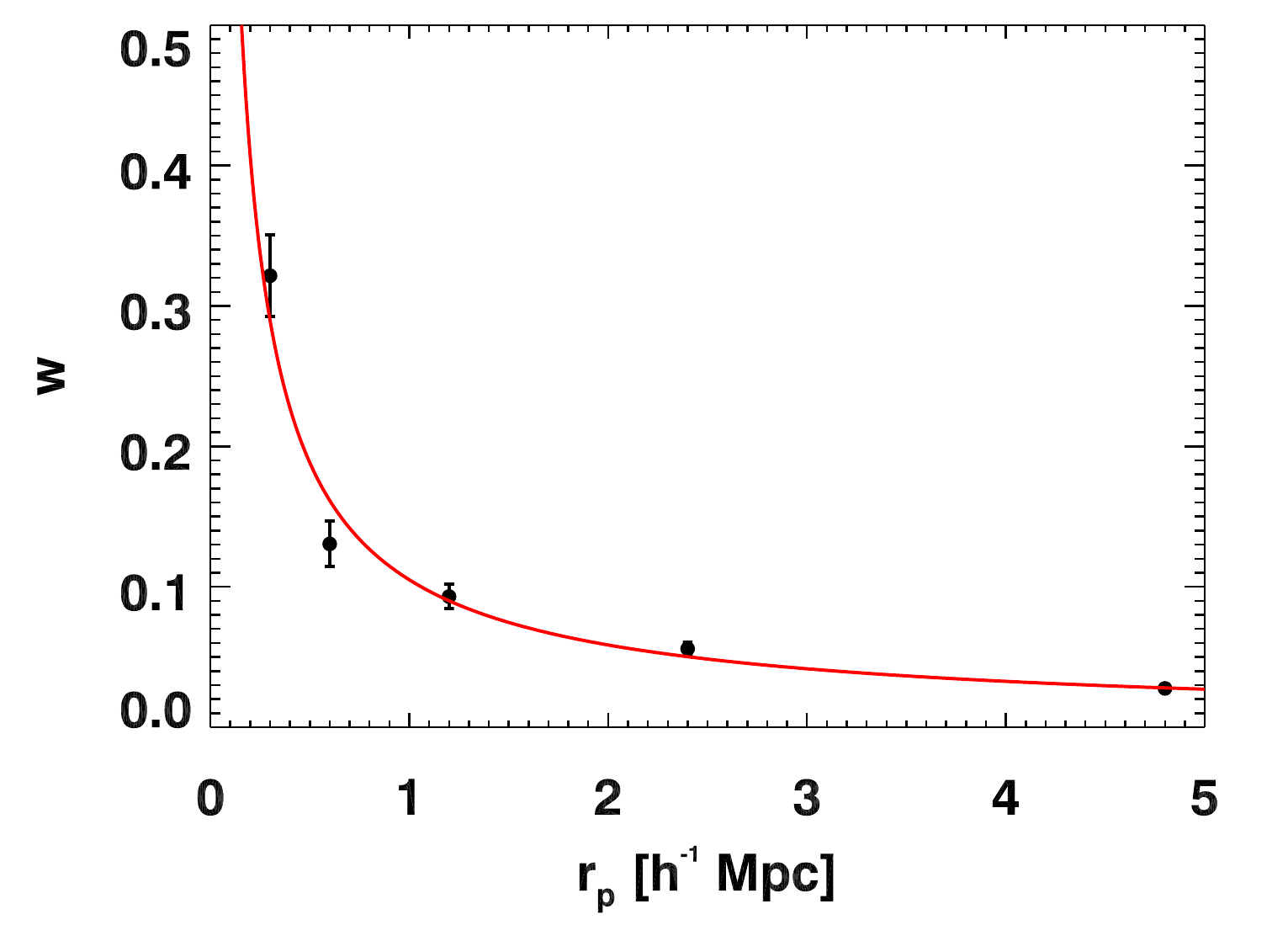,height=6.3cm,clip=} &
\hspace{-20pt}
\psfig{file=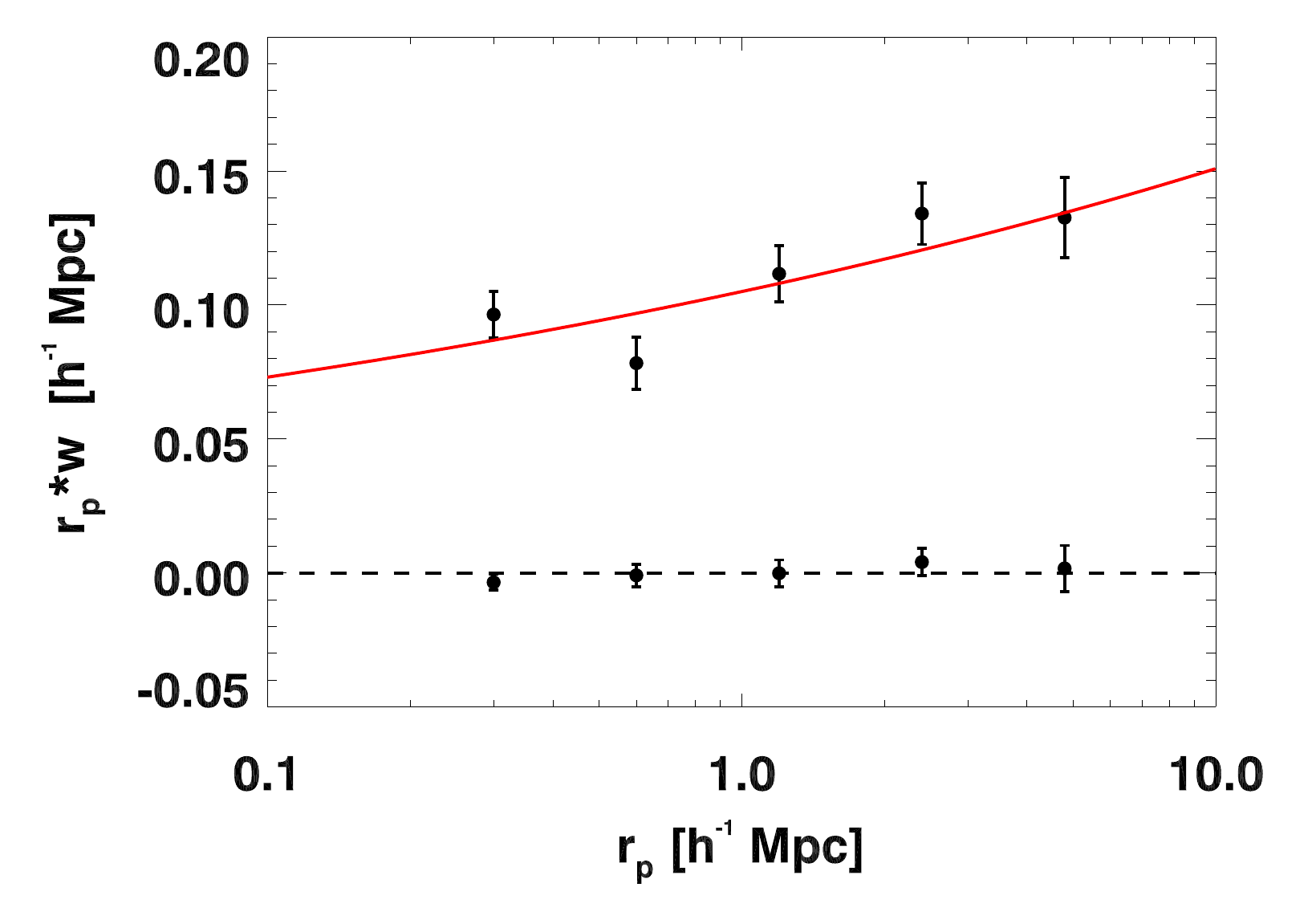,height=6.3 cm,clip=} \\
\hspace{-20pt}
\psfig{file=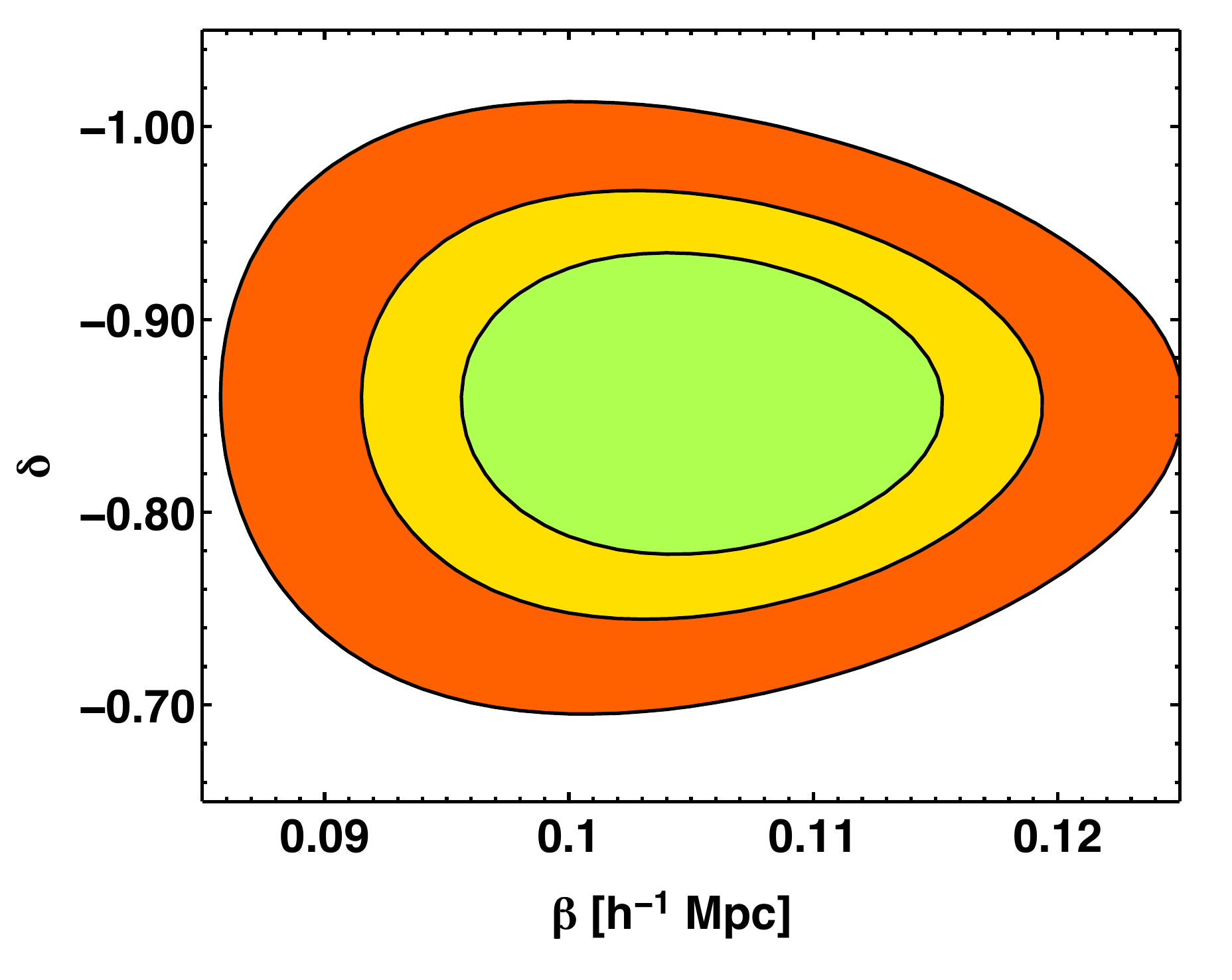,height=6.3 cm,clip=} &
\hspace{-20pt}
\psfig{file=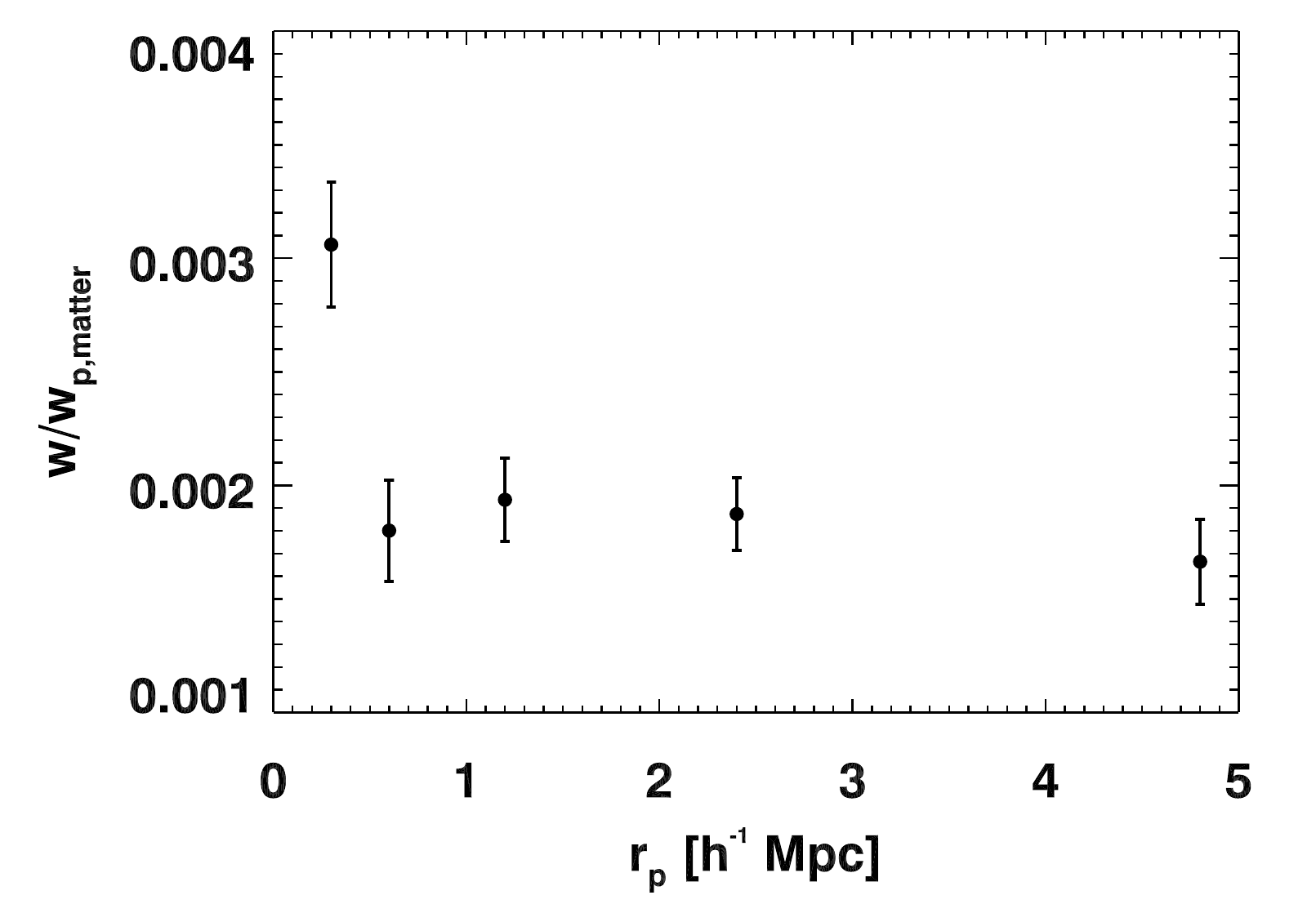,height=6.3 cm,clip=}
\end{tabular}
\caption[]{\small \textit{Top left}: Angular overdensity $w(r_p)$ and two-parameter curvefit $w = \beta r_p^{-\delta}$ for the cross-correlation of 4.2 million WISE galaxies 
about 7,049 quasars from DR7 and DR10.  Errorbars are 1 $\sigma$, and $r_p$ is measured in
comoving $h^{-1}$ Mpc.
\textit{Top right}: Same data as the top left panel, but with the $y$-axis replaced by $r_p w(r_p)$ to show
that we measure a significant overdensity of galaxies even at separations > 1 $h^{-1}$ Mpc.  The curvefit is shown in red.
The lower points are from the cross-correlation of the WISE galaxies with 37,402 $2.2 < z < 3.5$ quasars, measured in the same angular bins
as the $z \sim 0.8$
sample.
\textit{Bottom left}: $\chi^{2}$ contour for $\delta$ and $\beta$ with 68\%, 90\%, and 99\% confidence intervals.
\textit{Bottom right}: $w(r_p)$ divided by the projected correlation function of the linear-regime matter field, $w_{p,\textrm{matter}}(r_p)$, yielding a quantity proportional to the
quasar-galaxy bias $b_{\mathrm{QG}}^2$.
\label{fig:all_data}}
\end{figure*}

\subsection{Results}
\label{sec:results}

Figure~\ref{fig:all_data} shows the measured angular overdensity and a best-fit curve, using a sample of 7,049 quasars from both DR7 and DR10.  A two parameter fit
yields a minimum $\chi^{2}$ of 10.23 with 2 degrees of freedom, $\beta = 0.105 \pm 0.007$, and $\delta = 0.84 \pm 0.05$.  Varying the minimum bin radius does not substantially affect
$\beta$ or $\delta$, nor does successively removing each bin from the fit.

By dividing $w(r_p)$ by the projected correlation function of the linear-regime matter field,
we obtain a quantity proportional to $b_{\mathrm{QG}}^2$, the square of the linear quasar-galaxy bias (Equation~\ref{eqn:bias}).
We plot this quantity in Figure~\ref{fig:all_data}.
Quasar-galaxy clustering at 0.2--6.4 $h^{-1}$ Mpc arises from 
a mixture of one-halo and two-halo terms: the one-halo term refers to clustering within
the same dark matter halo, while the two-halo term refers to clustering between different
halos.  One-halo clustering leads to an increase in the linear bias at $r_p < 1$ $h^{-1}$ Mpc \citep{shen+13}
and cosmological hydrodynamic simulations indicate that the
one-halo term dominates clustering at $r_p \leq 0.3$ $h^{-1}$ Mpc \citep{deg+11}.
We interpret the sharp increase in linear bias at $r_p = 0.3$ $h^{-1}$ Mpc (Figure~\ref{fig:all_data})
as evidence for one-halo clustering in this bin.

\subsection{Testing for systematics}
\label{sec:systematics}

We check for systematic effects by measuring $w(r_p)$ for WISE galaxies about 37,402 quasars with $2.2 < z < 3.5$.  We randomly assign each quasar
a redshift $0.65 < z < 0.9$ so that we measure $w(r_p)$ on the same angular scales as for our $z \sim 0.8$ quasar sample.
We expect a small signal due to the gravitational lensing of high-redshift quasars by $z \sim 0.8$ galaxies.
The galaxies magnify the high-redshift quasars, lowering the flux limit in the region near the galaxy and creating
a cross-correlation \citep{my+03,my+05,scr+05,men+10}.  The strength of this signal is proportional to the magnification $\mu$:
\begin{equation}
w(r_p) \propto \mu \approx \frac{\theta_{\mathrm{E}}}{\theta}
\label{eqn:lensing}
\end{equation}
where $\theta_{\mathrm{E}}$ is the Einstein radius for a galaxy in our sample (typically $\approx 1"$ for galaxies at $z \approx 0.8$)
and $\theta$ is the angular separation between the galaxy and the quasar.
The strength of this signal depends on both the level of contamination by low-redshift galaxies and stars
and the average apparent magnitude of the quasar population.  
A linear least squares fit of constant $r_p w(r_p)$ yields $r_p w(r_p) = -0.0018 \pm 0.0023$.
However, for bins centered at 1000" and 2000" (9.6 and 19.2 $h^{-1}$ Mpc at $z = 0.8$, respectively),
we find a 2--3 $\sigma$ deviation from zero, with $w(r_p) \approx 0.003$ for these two bins.
Since the gravitational lensing signal peaks at much smaller angular scales,
we attribute this large-scale overdensity to a systematic error present on all scales.  While this error is very small compared to $w(r_p)$
at smaller scales, it is close enough
to $w(r_p)$ for the bins at $r_p$ = 9.6 and 19.2 $h^{-1}$ Mpc that we restrict our measurement of angular clustering to separations of 0.2--6.4 $h^{-1}$ Mpc.

\begin{figure}[H]
\psfig{file=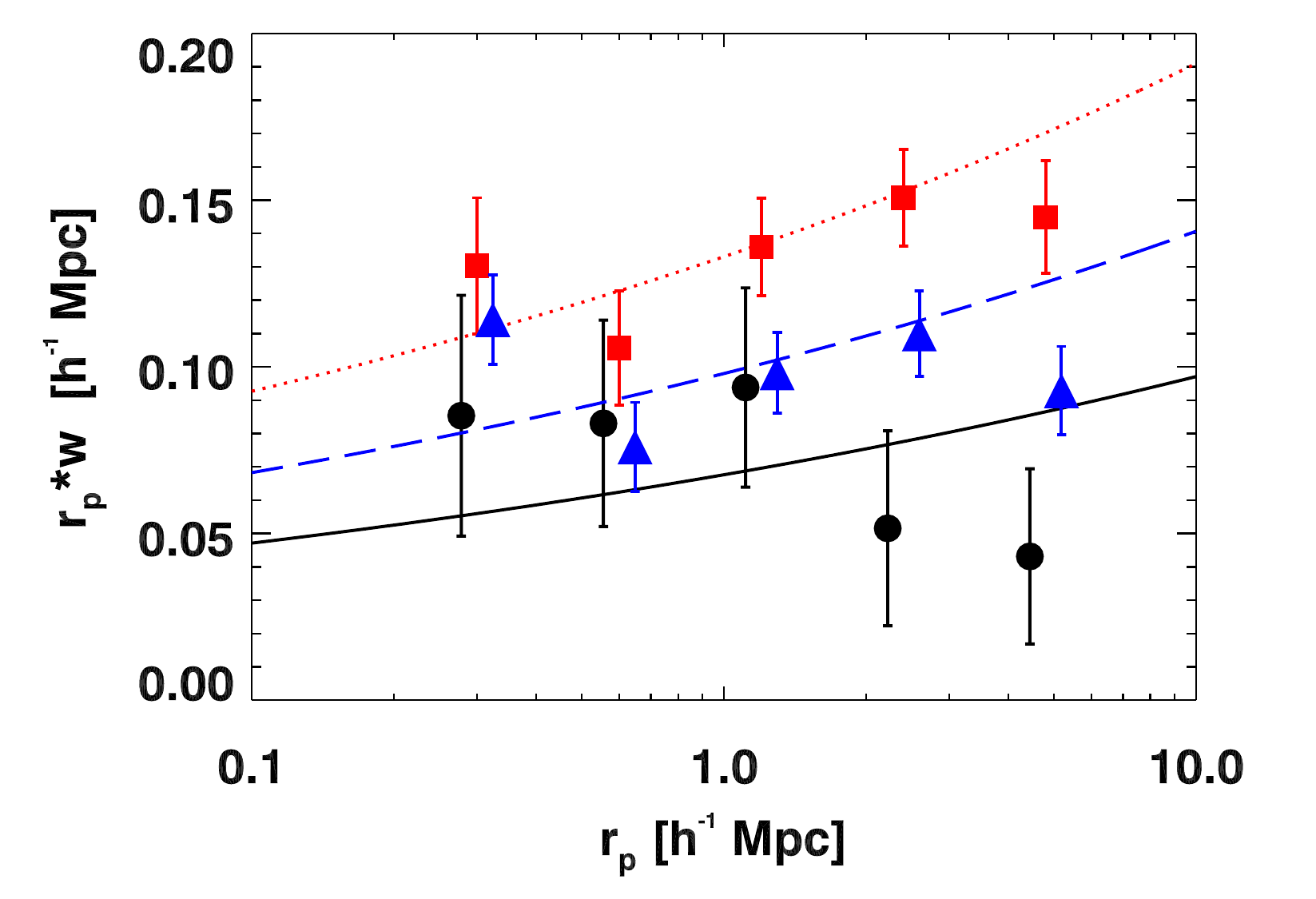,width=8 cm,clip=}
\caption[]{\small
Correlations between the DR7 and DR10 quasars and SDSS-identified extended sources (red squares, dotted line), non-matches (blue triangles, long-dashed line), and 
point sources (black circles, solid line).  The lines are linear least squares curvefits using $\delta = 0.84$, the slope from the two parameter fit in Figure~\ref{fig:all_data}.  Angular clustering
amplitudes at 1 $h^{-1}$ Mpc: extended sources, $\beta$ = 0.134 $\pm$ 0.009; non-matches, $\beta$ = 0.098 $\pm$ 0.008; point sources, $\beta$ = 0.068 $\pm$ 0.014.}
\label{fig:by_class}
\end{figure}

We separately cross-correlate each of the three components of the sample (SDSS-identified point sources, extended sources, and non-matches) 
with the full quasar sample and measure a clustering amplitude at 1 $h^{-1}$ Mpc at least 5 $\sigma$ greater than zero for all three cases (Figure~\ref{fig:by_class}).
The clustering amplitude for the SDSS point sources and the non-matches are 3 and 4 $\sigma$ lower than the clustering amplitudes for the SDSS extended
sources, respectively, indicating that the non-matches and point sources contain more contaminating stars and low redshift galaxies than the extended sources.
Nevertheless, all three components contain $0.65 < z < 0.9$ galaxies, and we believe it is prudent to include all three components in our sample to eliminate the possibility
of seeing-dependent variations in galaxy density.

\section{Dependence of Clustering Amplitude on Quasar Properties}
\label{sec:results}

We measure $w(r_p)$ for seven groups spanning 1.3 decades in luminosity and eight groups spanning 1.3 decades in virial mass.  
The seven groups in luminosity consist of three DR7 groups and four DR10 groups, while the eight groups in virial mass
consist of four DR7 groups and four DR10 groups (see Figure~\ref{fig:z_bh_luminosity} for luminosity and virial black hole mass distributions).

\begin{figure*}
\begin{tabular}{cc}
\psfig{file=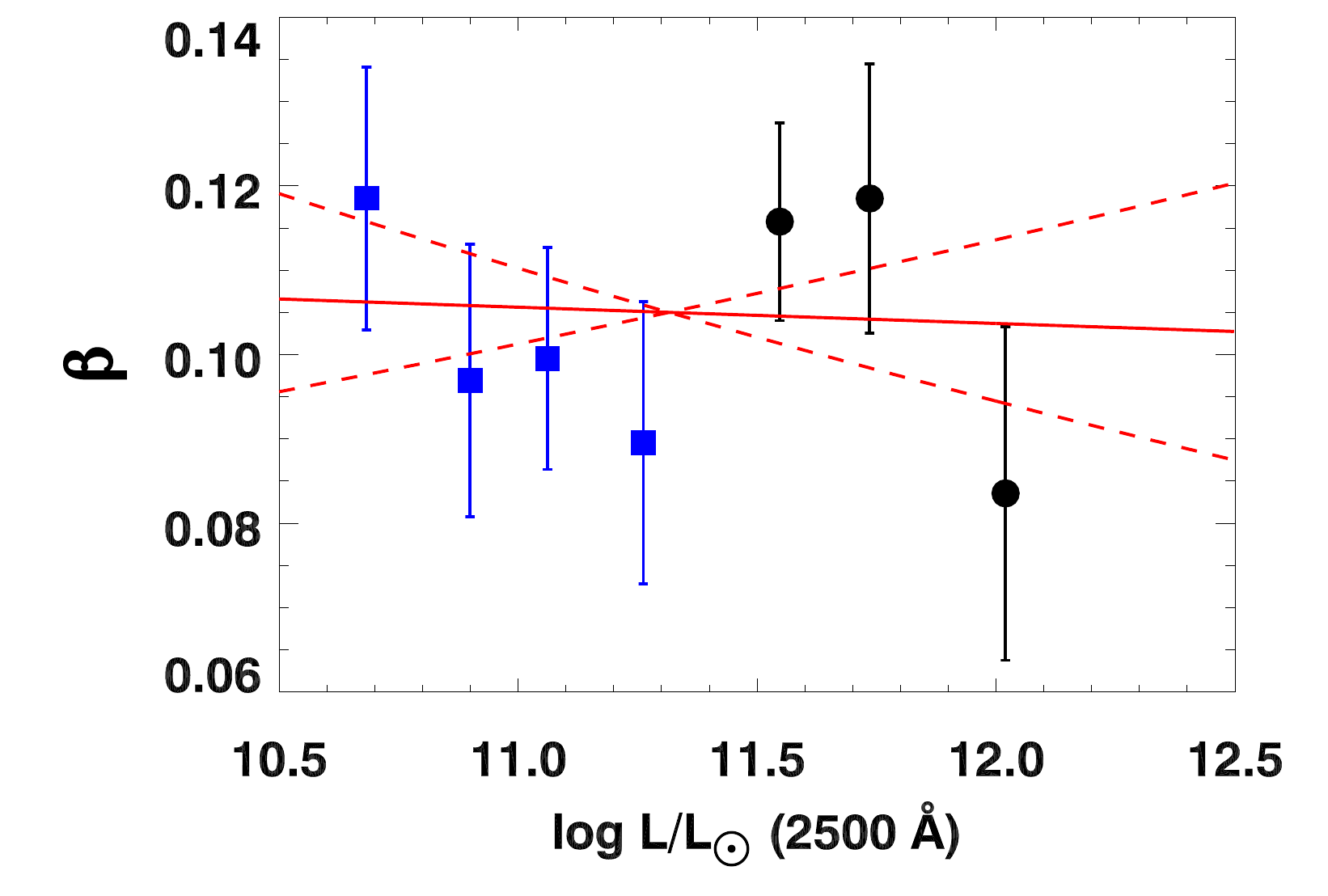,width=8 cm,clip=} &
\psfig{file=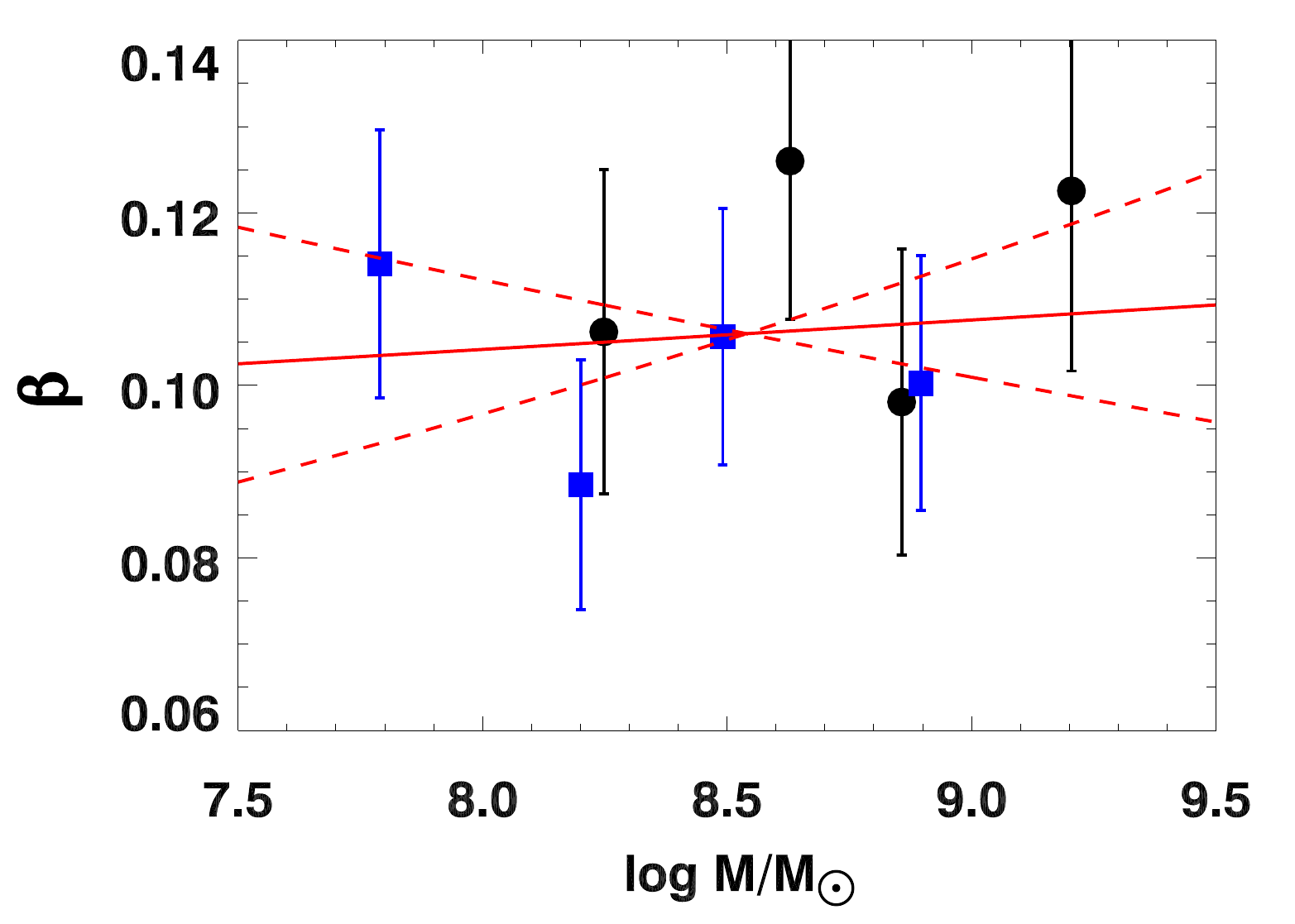,width=8 cm,clip=} \\
\psfig{file=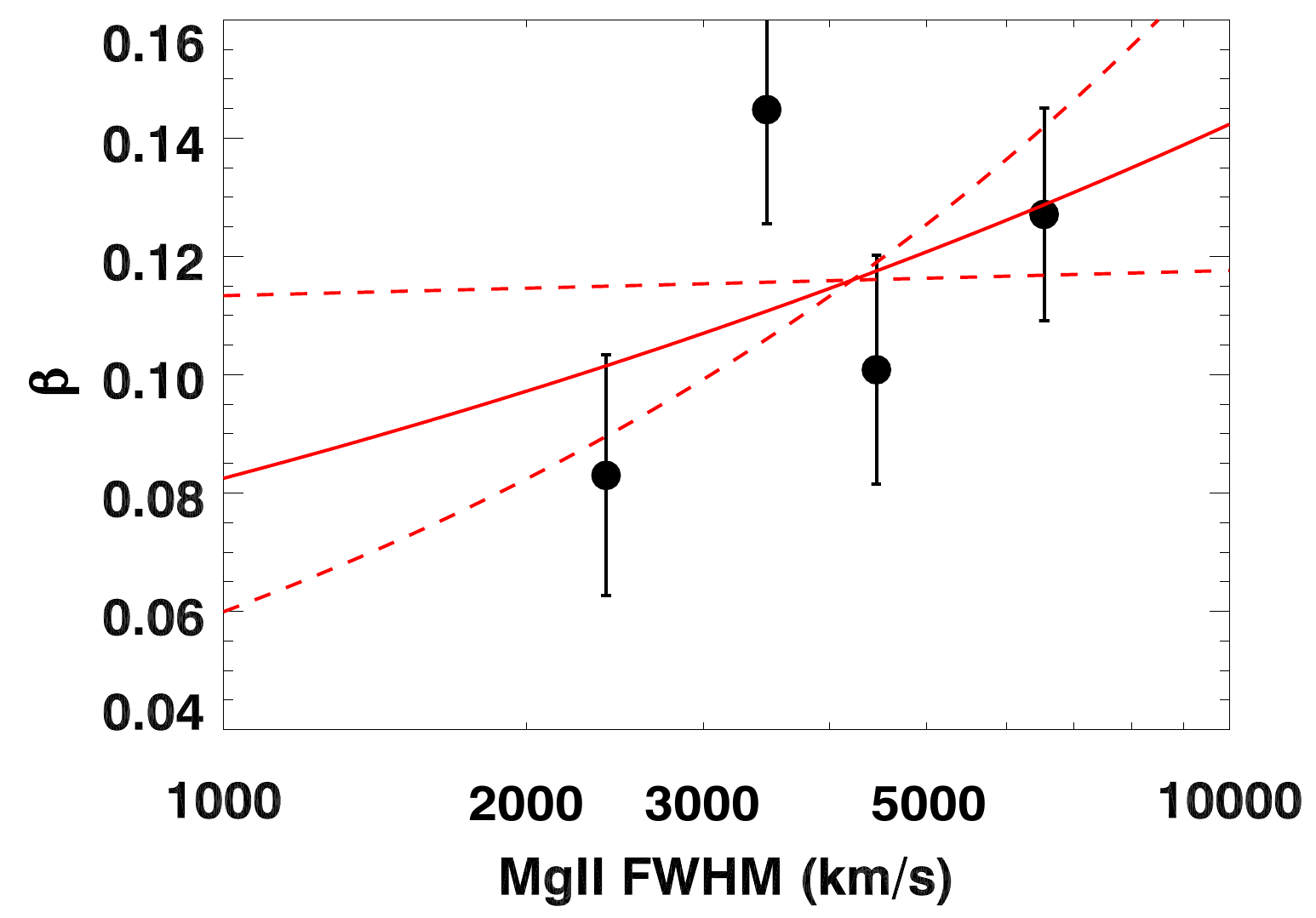,width=8 cm,clip=} &
\hspace{7pt}
\psfig{file=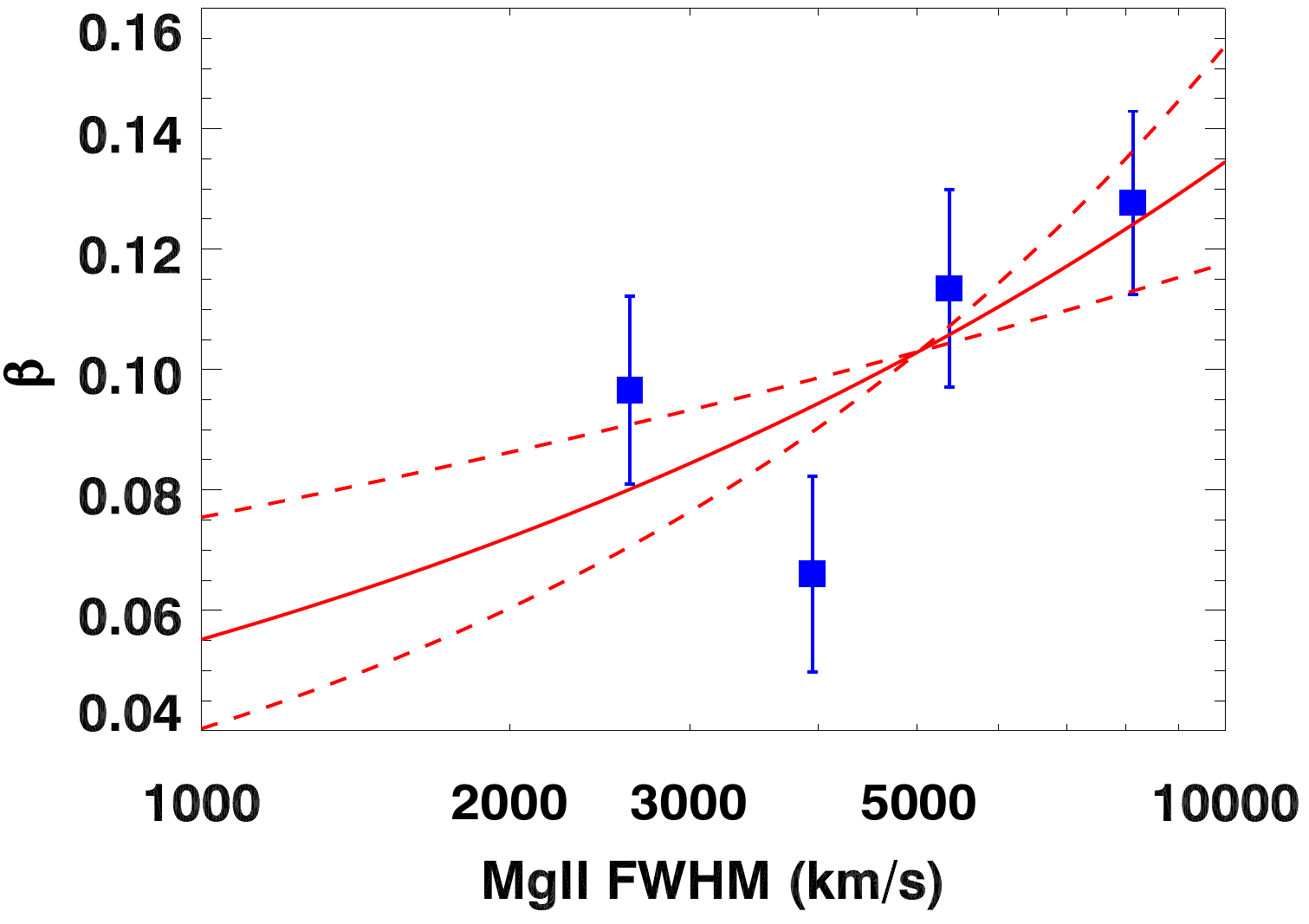,width=8 cm,clip=}
\end{tabular}
\caption[]{\small Luminosity dependence (\textit{top left}), virial black hole mass dependence (\textit{top right}),
and MgII FWHM dependence (\textit{bottom})
of $\beta$, the clustering amplitude at 1 $h^{-1}$ Mpc.  Luminosity dependence and virial mass dependence are presented
for both DR7 (black circles) and DR10 (blue squares) together, and FWHM dependence is presented separately for DR7 (\textit{bottom left})
and DR10 (\textit{bottom right}).
$\beta$ is determined from a linear least squares power-law fit to $w(r_p)$ measured in the five radial bins of Figure~\ref{fig:all_data} with $\delta$ fixed at 0.84.
Error bars are 1 $\sigma$.  The best-fit power law is shown in red.  The red dashed lines are power laws with the same amplitude
as the best-fit power law, but slopes $\pm 1 \sigma$ from the best fits.}
\label{fig:a_vs_bh}
\end{figure*}

We use linear least squares to find the clustering amplitude at 1 $h^{-1}$ Mpc ($\beta$) and its standard deviation for each group,
using all five radial bins in each fit.  Since we do not expect
the power-law slope to vary with luminosity or black  hole mass, we fix the slope at $\delta = 0.84$ in each fit.
If we allow the slope to vary, our results for $\beta$ differ by less than 0.3 $\sigma$ for all luminosity
and black hole mass subsamples.
Table~\ref{tab:all_fits} gives the luminosity, virial mass,
clustering amplitude and $\chi^{2}$ for each of these samples, Figure~\ref{fig:a_vs_bh} plots luminosity against clustering amplitude for the seven groups
in luminosity and  virial mass against clustering amplitude for the eight groups in virial mass.

We do not observe any luminosity dependence in the quasar-galaxy clustering amplitude.  We fit a power law between luminosity $L$ and clustering amplitude $\beta$:
\begin{equation}
\beta = a \left(\frac{L}{10^{11.32}} \right) ^{p}
\label{eqn:lum_curvefit}
\end{equation}
The factor in the denominator, $10^{11.32}$, is the geometric mean of the luminosities for each of the seven bins.
We find $p = \mathrm{d}\log{\beta} / \mathrm{d}\log{L} = -0.01 \pm 0.06$, $a = 0.105 \pm 0.006$ and $\chi^{2} = 4.75$ with 4 degrees of freedom.

We also find no relationship between quasar-galaxy clustering amplitude and virial black hole mass (Figure~\ref{fig:a_vs_bh}).
We fit a power law between virial mass $M$ and clustering amplitude $\beta$:
\begin{equation}
\beta = a \left( \frac{M}{10^{8.54}} \right) ^{p}
\label{eqn:bh_curvefit}
\end{equation}
As before, the factor in the denominator is the geometric mean of the virial masses of the eight bins.  Minimizing $\chi^{2}$ yields $p = 0.02 \pm 0.06$, $a = 0.106 \pm 0.006$
and $\chi^{2} = 3.88$ with 5 degrees of freedom.

\begin{deluxetable*}{lllllll}
\tablecolumns{7}
\tabcolsep0.1in\footnotesize
\tablewidth{0pc}
\tablecaption{
\label{tab:all_fits}}
\tablehead{
\colhead{Quasar Sample}       &
\colhead{Galaxy Sample}     &
\colhead{$N_{QSO}$}        &
\colhead{$\log{L/L_{\odot}}$} &
\colhead{$\log{M_{\textrm{BH}}/M_{\odot}}$} & 
\colhead{$\beta \pm 1 \sigma$} & 
\colhead{$\chi^{2}$}
}
\startdata
DR7 and DR10, $0.65 < z < 0.9$ & All & 7049 & 11.28 & 8.49 & 0.105 $\pm$ 0.006 & 10.23 \\
DR7 and DR10, $0.65 < z < 0.9$ & Point sources & 7049 & 11.28 & 8.49 & 0.068 $\pm$ 0.014 & 2.45 \\
DR7 and DR10, $0.65 < z < 0.9$ & Extended sources & 7049 &  11.28 & 8.49 & 0.133 $\pm$ 0.009 & 4.72 \\
DR7 and DR10, $0.65 < z < 0.9$ & Non-matches & 7049 & 11.28 & 8.49 & 0.098 $\pm$ 0.008 & 8.63 \\
DR10, $0.65 < z < 0.9$, $\log{L} < 10.81\textrm{ }\log{L_{\odot}}$ & All & 1052  & 10.68 & 8.10 & 0.119 $\pm$ 0.016 & 6.11 \\
DR10, $0.65 < z < 0.9$, $10.81\textrm{ }\log{L_{\odot}} < \log{L} < 10.98\textrm{ }\log{L_{\odot}}$ & All & 1051 & 10.90 & 8.29 & 0.097 $\pm$ 0.016 & 14.76 \\
DR10, $0.65 < z < 0.9$, $10.98\textrm{ }\log{L_{\odot}} < \log{L} < 11.15\textrm{ }\log{L_{\odot}}$ & All & 1052 &  11.06 & 8.41 & 0.100 $\pm$ 0.013 & 1.33 \\
DR10, $0.65 < z < 0.9$, $11.15\textrm{ }\log{L_{\odot}} < \log{L} < 11.43\textrm{ }\log{L_{\odot}}$ & All & 865 & 11.26 & 8.54 & 0.090 $\pm$ 0.017 & 4.49 \\
DR7, $0.65 < z < 0.9$, $\log{L} < 11.66\textrm{ }\log{L_{\odot}}$ & All & 1422 & 11.55 & 8.64 & 0.116 $\pm$ 0.012 & 1.16 \\
DR7, $0.65 < z < 0.9$, $11.66\textrm{ }\log{L_{\odot}} < \log{L} < 11.83\textrm{ }\log{L_{\odot}}$ & All & 711 & 11.74 & 8.71 & 0.119 $\pm$ 0.016 & 5.90 \\
DR7, $0.65 < z < 0.9$, $\log{L} > 11.83\textrm{ }\log{L_{\odot}}$ & All & 710 & 12.02 & 8.92 & 0.084 $\pm$ 0.020 & 5.72 \\
DR10, $0.65 < z < 0.9$, $\log{M_{\textrm{BH}}} < 8.04\textrm{ }\log{M_{\odot}}$ & All & 1047 & 10.88 & 7.79 & 0.114 $\pm$ 0.016 & 3.92 \\
DR10, $0.65 < z < 0.9$, $8.04\textrm{ }\log{M_{\odot}} < \log{M_{\textrm{BH}}} < 8.35\log{M_{\odot}}$ & All & 1047 & 10.96 & 8.20 & 0.088 $\pm$ 0.014 & 3.87 \\
DR10, $0.65 < z < 0.9$, $8.35\textrm{ }\log{M_{\odot}} < \log{M_{\textrm{BH}}} < 8.64\log{M_{\odot}}$ & All & 1047 & 11.01 & 8.49 & 0.106 $\pm$ 0.015 & 2.21 \\
DR10, $0.65 < z < 0.9$, $\log{M_{\textrm{BH}}} > 8.64\log{M_{\odot}}$ & All & 1048 & 11.11 & 8.90 & 0.100 $\pm$ 0.015 & 2.00 \\
DR7, $0.65 < z < 0.9$, $\log{M_{\textrm{BH}}} < 8.51\textrm{ }\log{M_{\odot}}$ & All & 703 & 11.62 & 8.25 & 0.106 $\pm$ 0.019 & 4.05 \\
DR7, $0.65 < z < 0.9$, $8.51\textrm{ }\log{M_{\odot}} < \log{M_{\textrm{BH}}} < 8.76\log{M_{\odot}}$ & All & 703 & 11.68 & 8.63 & 0.126 $\pm$ 0.018 & 2.12 \\
DR7, $0.65 < z < 0.9$, $8.76\textrm{ }\log{M_{\odot}} < \log{M_{\textrm{BH}}} < 8.99\log{M_{\odot}}$ & All & 703 & 11.71 & 8.86 & 0.098 $\pm$ 0.018 & 1.42 \\
DR7, $0.65 < z < 0.9$, $\log{M_{\textrm{BH}}} > 8.99\log{M_{\odot}}$ & All & 704 & 11.80 & 9.20 & 0.123 $\pm$ 0.020 & 8.74
\enddata
\tablecomments{Clustering amplitude ($\beta$), standard deviation, $\chi^{2}$, sample size, and
mean luminosity and virial black hole mass for the quasar subsamples.  The mean luminosity
and virial mass are weighted using the DR10 redshift distribution (\={z} = 0.77)
divided by the redshift distribution of each quasar subsample.
 The clustering amplitudes are derived from the measurements of $w(r_p)$ using two-parameter $\chi^2$ minimization
for the fit to the combined data (first row) and linear least squares assuming $\delta = 0.84$ for
all other subsamples.}
\end{deluxetable*}

Since the continuum luminosity is used to estimate virial black hole mass (Equation~\ref{eqn:virial_mass}),
$i$-band luminosity and virial black hole mass are correlated (correlation coefficient $\rho = 0.40$), and
the constraints on luminosity dependent clustering and black hole mass dependent clustering are not independent.
In order to provide an independent constraint on clustering, we also measure the dependence of clustering on MgII
FWHM.  If the luminosity (and therefore the broad-line region radius) is held constant, all of the variation in line width
can be attributed to variation in black hole mass.  We find that dividing the DR7 quasars into four equally sized groups
based on MgII FWHM produces four groups with very similar luminosity distributions; similarly, dividing the DR10 quasars
into four equally sized MgII FWHM groups produces four very similar luminosity distributions.  Since the DR7 and DR10
quasars have very different luminosity distributions, we compute the MgII FWHM dependence of clustering separately for
DR7 and DR10.  We also remove quasars with MgII rest-frame equivalent width < 10 \AA\  from our sample
(6 in DR7 and 13 in DR10) because \citet{shen+11} find that if the equivalent width is low, the FWHM of low S/N quasars
is biased low (see their Figure 7).
For DR7, we find $p = 0.24 \pm 0.22$, and for DR10, we find $p = 0.39 \pm 0.19$ for a power-law fit
between clustering amplitude and MgII FWHM.  While these results hint at a FWHM-dependent trend, more data
is needed to determine whether clustering is better correlated with FWHM than with virial mass.

Simulations indicate that the luminosity and black-hole mass dependence
of clustering is stronger for the one-halo term than for the two-halo term \citep{kh+02,thack+09}.  Since we interpret the sharp increase in bias
at $r_p = 0.3$ $h^{-1}$ Mpc (Figure~\ref{fig:all_data}) as indicative of one-halo clustering in this bin,
we measure the luminosity and black hole mass dependence of clustering using this bin only
to isolate the luminosity and black hole mass dependence of one-halo clustering.
We do not detect luminosity or virial mass dependence of clustering in the smallest bin:
for the luminosity dependence we find $p = 0.03 \pm 0.11$,
$a = 0.33 \pm 0.03$ and $\chi^2 = 7.16$ with 4 degrees of freedom, whereas 
for the virial mass dependence we find $p = 0.14 \pm 0.11$,
$a = 0.32 \pm 0.03$ and $\chi^2 = 8.54$ with 5 degrees of freedom.

To verify the robustness of our results, we repeat the measurements of luminosity and virial mass
dependent clustering using different samples
of quasars and galaxies.  We restrict the galaxy sample first by excluding SDSS-identified point sources
and then by also excluding non-matches.  
When we exclude point sources, we find identical slopes in the $\beta$-luminosity
and $\beta$-virial mass relationships.  When we exclude both point sources and non-matches,
we find that the slopes of the $\beta$-luminosity and $\beta$-virial mass relations are lower by about 2 $\sigma$.
This suggests that our results are not affected by systematic errors arising from the sample composition.

We also study the black hole mass dependence of clustering using virial masses from H$\beta$ instead MgII.
We use a restricted redshift range, $0.65 < z < 0.85$, to ensure that the entire H$\beta$ profile is measured
by the DR7 spectrograph.  We obtain a slope of $p = 0.07 \pm 0.07$ for the H$\beta$ masses compared to
$p = 0.00 \pm 0.06$ for the MgII masses in this range.  Thus, we detect no dependence of clustering amplitude
on either H$\beta$ or MgII based virial masses.

As discussed in Section~\ref{sec:quas_selection}, the virial black hole masses possess significant uncertainties of up to 0.5 decades \citep{shen_mass+13}.
As a result, the mean true mass of each bin is less extreme than the mean virial mass, suppressing the virial mass dependence
of quasar-galaxy clustering.
We estimate this suppression by simulating our data using an assumed power-law relationship between $\beta$ and true mass,
scattering true mass to virial mass, binning the simulated data by virial mass, and fitting a power law to the binned data.
We then find the ratio between the assumed power law slope and the fitted slope.
To find the distribution of true black hole masses, we approximate our observed virial mass distribution as lognormal with scatter
$\sigma_{\mathrm{obs}}$ (0.38 decades for DR7 and 0.43 decades for DR10) 
and assume that the distribution of virial masses at fixed true mass is lognormal with scatter $\sigma_{\mathrm{vir}}$; then the true mass distribution
is lognormal with scatter $\sigma_{\mathrm{true}} = \sqrt{\sigma_{\mathrm{obs}}^{2}-\sigma_{\mathrm{vir}}^{2}}$.
Previous measurements of the black hole mass function,
using virial masses but explicitly modelling both the luminosity-dependent bias and the Malmquist bias, found $\sigma_{\mathrm{true}} = 0.26$
and $\sigma_{\mathrm{vir}} = 0.26$ for DR7 quasars at $z = 0.8$ \citep{sk+12}.

We compute the ratio between the assumed power-law slope and the measured power-law slope for different values of $\sigma_{\mathrm{vir}}$, finding
that the ratio reaches 2 at $\sigma_{\mathrm{vir}} = 0.33$ ($\sigma_{\mathrm{true}} = 0.19$ for DR7, 0.28 for DR10) and asymptotes as $\sigma_{\mathrm{vir}}$ approaches $\sigma_{\mathrm{obs}}$.  
We also note that if luminosity-dependent bias is present, $\sigma_{\mathrm{true}} > \sqrt{\sigma_{\mathrm{obs}}^2-\sigma_{\mathrm{vir}}^2}$, since the luminosity-dependent bias
will decrease the observed scatter in the virial masses.  Since \citet{shen_mass+13} and \citet{sk+12} find evidence for luminosity-dependent bias
for MgII-based masses at $z = 0.8$, $\sigma_{\mathrm{vir}} = 0.3-0.35$ is most consistent with previous 
measurements of the black hole mass function and estimates of $\sigma_{\mathrm{vir}}$.  
This value of $\sigma_{\mathrm{vir}}$ corresponds to a suppression of the power-law slope
by a factor of 1.5--2.

\section{Discussion}
\label{sec:disc}

\subsection{Slope of the correlation function}
\label{sec:slope}

Comparing the clustering amplitude $\beta$ to previous measurements of the correlation length $r_{0}$
is beyond the scope of this paper, since it would require converting our measurements of the angular overdensity to measurements
of the projected cross-correlation function using the three-dimensional density of WISE galaxies $\rho_{0}$ (Equation~\ref{eqn:wtoomega}).
However, because of the scale independence of the power law fit, we can compare our power-law slope,
 $\delta = 0.84 \pm 0.05$, to other results, using Equation~\ref{eqn:proj_ccf} to relate $\delta$ to other measurements
 of $\gamma = \delta + 1$.
\citet{c+07} finds $\gamma$ between 1.6 and 2 depending on the sample, with 1 $\sigma$ errorbars $\sim$ 0.2.
\citet{por+06} find $\gamma = 1.7^{+0.53}_{-0.67}$ for quasars and galaxies with redshift $0.8 < z < 1.06$, \citet{shen+13} finds $\gamma = 1.69 \pm 0.07$ for redshifts $0.3 < z < 0.9$,
and \citet{zh+13} finds $\gamma = 2.1 \pm \sim 0.2$ for redshifts $0.6 < z < 1.2$.

Our measurement of $\delta$ is similar to most of these results, although very slightly higher.
A slightly steeper correlation function is to be expected for our sample: our color cut biases the galaxy
sample in favor of red galaxies, which cluster more strongly than blue galaxies.
According to the galaxy autocorrelation function measurements of \citet{zeh+11}, red galaxies have $\gamma = 1.94 \pm 0.03$, whereas blue galaxies have $\gamma = 1.66 \pm 0.03$.
Furthermore, we may measure a steeper slope because we measure quasar-galaxy clustering at smaller scales than previous studies.
While we measure clustering from 0.2--6.4 $h^{-1}$ Mpc,
\citet{c+07} measure clustering from 0.1--10 $h^{-1}$ Mpc, \citet{por+06} measure clustering from 3--20
$h^{-1}$ Mpc, and \citet{shen+13} measure clustering from 2--25 $h^{-1}$ Mpc.
Therefore, our measurements are more sensitive to the one-halo term than previous measurements, leading
to a steeper power law slope at scales < 1 $h^{-1}$ Mpc \citep[e.g.][]{shen+13}.  Indeed, our results suggest a break in the power law at 1 $h^{-1}$ Mpc,
with a steeper slope at smaller scales and a shallower slope at larger scales (see Figure~\ref{fig:all_data}).  However, we do not have enough bins to 
fit either a full halo occupation distribution (HOD) function or a broken power law.

\subsection{Luminosity dependent clustering: comparison to previous results}
\label{sec:lum_dependent}

We find no dependence of clustering amplitude upon luminosity (Figure~\ref{fig:a_vs_bh}).
We measure the luminosity dependence of clustering across a larger luminosity range and with more bins in luminosity than
previous studies at similar redshifts.  As a result, our power-law fit of clustering amplitude versus luminosity
provides a similar or tighter constraint on the luminosity dependence of quasar clustering than previous results.

Most previous studies of luminosity dependent clustering measure the correlation length $r_0$ as a function of luminosity.
Since $r_0 \propto \beta^{1/\gamma} = \beta^{0.55}$ (Equation~\ref{eqn:proj_ccf}), 
we fit a power law between luminosity and $\beta^{0.55}$ to compare our results to previous results, using errors
computed from Taylor series error propagation.  We find a slope of $0.002 \pm 0.027$.  
A power law is advantageous because it is scale independent, allowing us to compare 
our slope to the slopes obtained
from power-law fits between $r_0$ and luminosity.  When fitting power laws to other results,
we assume uncorrelated errors and use the same nonlinear $\chi^2$ minimization used to determine the amplitude and slope of the power law
fit to our data.  We compare to measurements of $r_0$ rather than to measurements of the quasar bias $b_Q$ because measurements
of the quasar bias
incorporate errors arising from measurement of the galaxy bias.

\citet{zh+13}, measuring the cross-correlation of SDSS Stripe 82 galaxies and SDSS quasars
at $z \sim 0.8$, found no significant differences in clustering between two
bins in luminosity with 10--25\% error on the clustering measurements.
\citet{zh+13} compared the clustering of faint and bright quasars for three different redshift bins,
yielding six bins in luminosity but with differing redshift distributions.  
In order to fit a power law to the correlation length measurements of \citet{zh+13},
we normalize their correlation length measurements in the $0.8 < z < 1.0$ and $1.0 < z < 1.2$ bins
so that the average correlation length in each redshift bin is identical.
We obtain a power law slope of 
$-0.01 \pm 0.10$ for the \citet{zh+13} measurements, for quasars at  \={z} = 0.9 and $10.98 < \log{L_{\odot}} < 11.57$.

\citet{sha+11} compared correlation lengths measured by SDSS, 2QZ, and 2SLAQ at $z = 1.4$
across slightly more than a decade in luminosity ($11.44 < \log{L_{\odot}} < 12.48$).
A power law fit to their results yields a slope of $0.04 \pm 0.03$, providing a similar constraint on the magnitude of luminosity-dependent
clustering as our result, although the measurements of \citet{sha+11} are at higher redshift than our results.

\citet{shen+13}, measuring the cross-correlation of DR10 CMASS galaxies and DR7 quasars
at $0.3 < z < 0.9$ (\={z} = 0.53) with  $11.19 < \log{L_{\odot}} < 11.88$,
found no luminosity dependence over four bins in luminosity.  We fit a power law to the
\citet{shen+13} measurements of luminosity and correlation length, obtaining an exponent $p = -0.01 \pm 0.05$.
While the clustering measurements in each bin of \citet{shen+13} have $\approx 10\%$ errors in $\xi(r)$,
compared to our errors of $\approx 15\%$ in $w(r_p)$, we provide a tighter constraint on the luminosity
dependence of clustering because we measure clustering across a larger luminosity range and with more bins in luminosity.

Comparison of our work to previous work shows that measuring clustering across a wide range of luminosity
is important in obtaining a more precise measurement of (or constraint on) luminosity-dependent clustering.
The restricted range in optical luminosities is often cited as a justification for the lack of measured luminosity
dependence \citep[e.g.][]{bon+09} and the wider luminosity range probed by X-ray selected AGN may explain why
X-ray luminosity dependent clustering has been detected while optical luminosity dependent clustering has not \citep{krum+12}.

\subsection{Luminosity dependent clustering: comparison to models}
\label{sec:lum_dependent_models}

\begin{figure}
\psfig{file=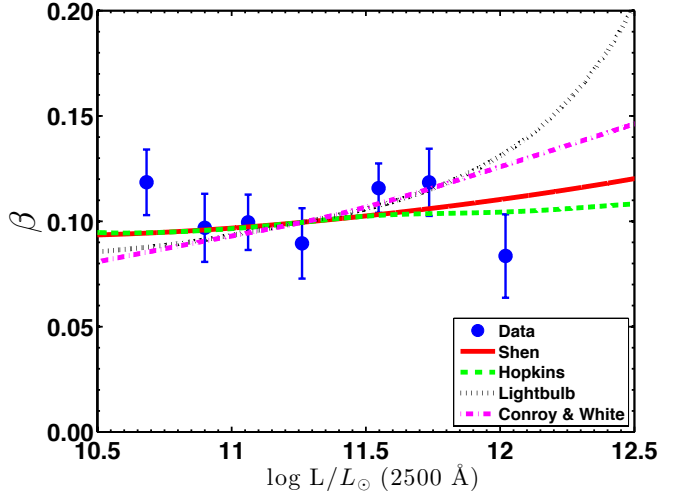,width=9 cm,clip=}
\caption[]{\small Measured clustering amplitude as a function of luminosity, compared to theoretical
predictions.  All four theoretical models predicted quasar bias as a function of luminosity.  We assumed
that quasar bias is proportional to clustering amplitude and normalized the relationship to the correlation
amplitude of the median luminosity bin.  Luminosities are measured using $M_{i}(z=2)$,
measured in solar luminosities, corresponding to rest-frame 2500 \AA\ luminosity.
\label{fig:compare_to_theory}}
\end{figure}

In Figure~\ref{fig:compare_to_theory}, we compare our results to theoretical predictions of luminosity-dependent clustering from four different models
of quasar evolution: \citet{shen_mod+09}, referred to as S09, \citet{hop+14}
(H14), \citet{cw+13} (CW13), and a simple ``lightbulb'' model used by \citet{hop+07} for comparison
to more sophisticated models.  
The S09 and H14 models are physically motivated, while the CW13 and lightbulb models 
merely provide a relationship between halo mass and quasar luminosity.

The ``lightbulb'' model is the simplest of the four considered.  We use the relationship given at $z = 1$ in Figure 3 in \citet{hop+07} (the line labelled
``All QSO at L = L$_{\textrm{Edd}}$'').
In this model, all quasars radiate at the same Eddington ratio and black hole mass is related to host halo mass without scatter.
Thus, the luminosity-bias relationship is entirely determined by the halo mass-bias relationship.  
However, observation of scatter in both the Eddington ratio distribution \citep{shen+08,koll+06} and in the halo mass-black hole mass relationship \citep{ferr+02}
indicate that the lightbulb model is an oversimplification.
The CW13 model (using the $z = 1$ relationship from Figure 8) is a ``scattered lightbulb'' model with scatter in both the Eddington ratio distribution
and in the black hole mass-host halo mass relation.  Similar to the lightbulb model, the CW13 model uses a step function for the quasar light curve and
uses the quasar lifetime to set the quasar bias at a given redshift.

In contrast, both the S09 and H14 models are based on physically motivated light curves.
The H14 model (using the $z = 1.5$ bias-luminosity relationship in their Figure 5) considers two modes of quasar fueling: major merger induced fueling, which is responsible for most high luminosity
quasars, and stochastic fueling mechanisms such as minor mergers,
which are responsible for lower luminosity quasars.  The merger-induced fueling model is quite similar to \citet{hop+07}, and predicts from merger simulations
that host halo mass and peak quasar luminosity are tightly linked \citep{lidz+06}.  However, the relationship between peak luminosity and instantaneous luminosity
has considerable scatter, arising from the merger-based quasar light curve of \citet{hop+05}.
H14 also considers stochastic accretion from low-luminosity quasars and Seyfert galaxies, which reside in disk hosts rather than bulges, although
the fraction of disk-hosted quasars at the luminosities of our sample is estimated to be quite small.
In the H14 model,
quasars spend most of their lives accreting at low rates.  As a result, many faint quasars have high peak luminosities and thus high host halo masses.
Thus, a large range in instantaneous luminosity maps to a small range in peak luminosity (and thus black hole mass, halo mass, and clustering strength).

The S09 model (using the relationship at $z = 1$ from their Figure 5) is similar to H14, differing in only three respects.  First, S09 estimates the major merger rate using simulations while H14 estimates
the merger rate from empirical halo distribution models.  Second, H14 derive their quasar light curve from simulations of galaxy mergers, while S09
fits the light curve empirically from observations.  Last, S09 does not consider alternative fueling mechanisms for low-luminosity quasars, but, as discussed in the paragraph above,
the practical significance of this choice is small.

Each model specifies the relationship between the linear quasar bias $b_{Q}$ and the quasar luminosity.  
In analogy with Equation~\ref{eqn:bias}, the linear bias $b_{Q}$ is defined using the quasar-galaxy cross-correlation function:
\begin{equation}
b_{Q} = \frac{\xi_{\mathrm{QG}}}{b_{G} \xi_{\textrm{matter}}}
\label{eqn:qg_bias}
\end{equation}
If a linear quasar bias provides an appropriate fit to our data, then Equations~\ref{eqn:ccf},~\ref{eqn:proj_ccf}, and~\ref{eqn:wtoomega}
imply that $\beta \propto b_{Q}$.  To compare our results to models predicting the luminosity dependence of quasar bias, we convert
$b_{Q}$ to $w(r_p)$ by
forcing each model to match the median $\beta$ at the median luminosity of our data.

The linear quasar bias is only applicable on larger scales.  At scales < 1 $h^{-1}$ Mpc, the one-halo term leads to nonlinear clustering
and deviations from Equation~\ref{eqn:qg_bias}, as shown in Figure~\ref{fig:all_data}.  However, Figure~\ref{fig:all_data} implies that a linear
bias is appropriate in all bins except the innermost bin, implying that $\beta \propto b_{Q}$ is not an unreasonable assumption.
Moreover, we expect that one-halo clustering will also depend on luminosity, perhaps more strongly than two-halo clustering \citep{kh+02,thack+09}.

The models present luminosity in different units.  The S09 model gives the relationship between bias and bolometric luminosity in erg s$^{-1}$.
We convert to $M_{i}(z=2)$ using Equation 1 from \citet{shen+09} and then to solar luminosity using the same methods as in this paper.
The H14 and lightbulb models express luminosity in terms of solar bolometric luminosity; we convert to erg s$^{-1}$ using the bolometric
luminosity of the sun, $L_{\odot} = 3.827 \times 10^{33}$ erg s$^{-1}$.  As before, we then use Equation 1 from \citet{shen+09} to convert
to $M_{i}(z=2)$ and the same relationship used in this paper to convert to solar luminosity.  The CW13 model gives luminosity
using $M_{i}(z=2)$, which we again convert to solar luminosity following the methods used above.

The H14 model provides the best fit to our data ($\chi^{2} = 5.96$ for 6 degrees of freedom), followed by the S09 model ($\chi^2 = 6.43$), the CW13 model ($\chi^2 = 10.38$),
and the lightbulb model ($\chi^{2} = 11.22$) (Figure~\ref{fig:compare_to_theory}).  
Our data prefer the H14 model to the lightbulb model at 2 $\sigma$.  However, at present none of these models
are disfavored by our data at more than 2 $\sigma$.  
The divergence between models is greatest at high luminosities, suggesting that improved constraints could be obtained by measuring
clustering at higher redshifts, or by using a deeper sample of tracer objects to obtain more precise clustering measurements.

\subsection{Black hole mass dependent clustering}
\label{sec:bh_dependent_clustering}

We find no significant relationship between virial black hole mass and clustering strength, in contrast to previous measurements
of black hole mass dependent clustering by \citet{fine+06} and \citet{komi+13}.  Our methods are quite
different from the methods of \citet{fine+06}; they measured the relationship between black hole mass and host halo mass for bins
at different redshift, allowing them to probe a larger dynamic range in black hole mass.  Moreover, \citet{fine+06} measured only a weak
relationship between black hole mass and host halo mass.  Our results are in more significant tension with those of \citet{komi+13}:
a power-law fit to the black hole mass dependence of the correlation length
$r_0$ from the data of \citet{komi+13} yields a slope of $0.14 \pm 0.05$, compared to a slope of $0.01 \pm 0.03$ for our data (fitting the black hole mass dependence
of $\beta^{0.55} \propto r_0$).  There are significant methodological differences between our results and those of \citet{komi+13}:
\citet{komi+13} measured black hole mass over a significantly broader redshift range ($0.1 <  z < 1.0$), used masses from three virial mass indicators
(MgII, H$\beta$, and H$\alpha$), and did not correct for differing redshift distributions between different bins in black hole mass,
although they found only minor differences using reselected bins containing identical redshift distributions.
We also note that other studies failed to find a significant relationship between black hole mass and clustering strength: 
\citet{shen+09}
found no dependence of quasar clustering on virial mass using two bins in viral mass at $0.4 < z < 2.5$,
and \citet{zh+13} found a 1--2 $\sigma$ difference in clustering strength between two bins in virial mass at $0.6 < z < 1.2$.
Furthermore, \citet{cha+13} found no significant relationship between host cluster mass and virial mass at $0.1 < z < 0.3$;
rather than using clustering to measure the linear bias, they directly matched quasars with galaxy clusters
and estimated the cluster mass using the optical richness.

Our results also appear to conflict with the predictions of many
different models, which predict a relationship between black hole mass and host halo mass \citep{lidz+06,kh+02,cw+13,ferr+02}.
These models
all predict $M_{\mathrm{BH}} \propto M_{\mathrm{DMH}} ^{\alpha}$ over the range of masses considered here, with 1.2 < $\alpha$ < 1.8.
We compare our data to these predictions by estimating the $\beta$-black hole mass
relationship that would result from these $M_{\mathrm{BH}}-M_{\mathrm{DMH}}$ relationships.

We begin by relating $M_{\mathrm{DMH}}$ to the linear quasar bias $b_{Q}$.  We use the fitting formula of \citet{she+01} to find $b(M)$:
\begin{equation}
\begin{split}
b = 1 + \frac{1}{\sqrt{a} \delta_{sc}(z)}  \bigg[\sqrt{a} (a \nu^2) + \sqrt{a} b (a\nu^2)^{1-c} \\
- \frac{(a\nu^2)^c}{(a\nu^2)^c + b(1-c)(1-c/2)} \bigg]
\end{split}
\label{eqn:sheth_bias}
\end{equation}
where $a = 0.707$, $b = 0.5$, $c = 0.6$, $\nu = \delta_{sc}(z)/\sigma(M,z)$ and $\delta_{sc}(z) = 0.15(12\pi)^{2/3} \Omega_{m}^{0.0055}$
\citep{nfw+97}.
$\sigma(M,z) = \sigma(M,z=0) D(z)$, where $D(z)$ is the linear growth factor.
We approximate $D(z)$ using the fitting form of \citet{car+92}:
\begin{equation}
\begin{split}
& D(z) = \frac{D_1(z)}{D_1(0)} \\
& D_1(z) = \left(\frac{1}{1+z}\right) \frac{5}{2} \Omega_{mz} \bigg[ \Omega_{mz}^{4/7} - \Omega_{\Lambda z} + \left(1 + \frac{\Omega_{mz}}{2}\right) \\
& \left(1+\frac{\Omega_{\Lambda z}}{70}\right)\bigg]^{-1}
\label{eqn:growth}
\end{split}
\end{equation}
where $\Omega_{mz}$ and $\Omega_{\Lambda z}$ give the evolution of the cosmological parameters $\Omega_{m}$ and $\Omega_{\Lambda}$.
$\sigma(M,z=0)$ is the fluctuation in the density field for a halo of mass $M$ at redshift 0, given by
\begin{equation}
\sigma^2(r) = \frac{1}{2 \pi^2}\int_{0}^{\infty} \! k^2 P(k) W^2(kr) \, \mathrm{d}k
\label{eqn:sigma}
\end{equation}
where $P(k)$ is the power spectrum of the linear-regime matter field, computed from the transfer function of \citet{eh+98},
\begin{equation}
W(kr) = \frac{3[\sin{kr} - kr\cos{kr}]}{(kr)^3}
\label{eqn:omega}
\end{equation}
and $r(M) = (3M / (4 \pi \rho_0))^{1/3}$, where $\rho_0$ is the mean density of the Universe at $z = 0$, 2.78 $\times$ 10$^{11} \Omega_m h^2 M_{\odot}$ Mpc$^{-3}$.
We normalize the linear power spectrum such that $\sigma(r = 8 h^{-1} Mpc) = 0.84$.

Since our angular clustering measurement does not allow for accurate estimates of $b_Q$,
we relate $b_{Q}$ to $\beta$ using previous measurements of $b_{Q}$ at $z = 0.8$.  \citet{cro+05} measured $b_{Q}(z = 0.8) = 1.49 \pm 0.21$;
they also provided an empirical quadratic fit to their measurements of $b_{Q}$ from $z = 0.53$ to $z = 2.48$, yielding $b_{Q}(z = 0.8) = 1.47 \pm 0.2$.
We fit a quadratic function to the measurements of \citet{shen+09,shen+13,whi+12,pad+09,cro+05,por+06}, using 29 different measurements with $0.3 < z < 3.8$
and obtain $b_{Q}(z = 0.8) = 1.43$.  Following these estimates, we use $b_{Q} = 1.5$ as the linear bias of our sample.  Since $b_{Q} \propto \beta$, we multiply $b(M)$ by the ratio
between $\beta = 0.105$ and $b(z=0.8) = 1.5$.  Since $b(M)$ is approximately linear over 1 decade in host halo mass, $\beta(M_{\mathrm{BH}})$ is well-approximated by a power law.

The power-law slopes $1.2 < \mathrm{d} \log{M_{\mathrm{BH}}} / \mathrm{d} \log{M_{\mathrm{DMH}}} < 1.8$ correspond to $0.1 < \mathrm{d} \log{\beta} / \mathrm{d} \log{M_{\mathrm{BH}}} < 0.15$.  We only find $\approx 1 \sigma$
difference between these slopes and our measurement of the MgII virial mass dependence of clustering, after adjusting for the factor of 1.5--2 suppression arising
from the large scatter in the virial masses.  
Our results are also consistent with no relationship between $M_{\mathrm{BH}}$ and $M_{\mathrm{DMH}}$, a possibility suggested by \citet{sha+11},
who argue that the lack of luminosity-dependent clustering and the corresponding long quasar lifetime supports a pure luminosity evolution model,
in which bright quasars at high redshift become faint Seyfert galaxies at $z = 0$.  In the pure luminosity evolution model,
host halo mass is not tightly related to black hole mass, as supported by the results of \citet{kor+11} and \citet{kb+11}, who found no relationship
between black hole mass and stellar disk mass in disk galaxies at $z = 0$.  Our results suggest that discriminating between models with no link
between black hole mass and host halo mass and models with a tight relationship between black hole mass and host halo mass
will require clustering measurements with larger samples and larger dynamic range in black hole mass.

\section{Conclusions}
\label{sec:conclusions}

Measuring the luminosity dependence of the quasar clustering amplitude allows us to test different relationships between host halo mass
and quasar luminosity.
Previous measurements of the quasar autocorrelation function and the three-dimensional
quasar-galaxy cross-correlation function have suffered from small sample sizes: the spatial density of quasars is low, and it 
is difficult to obtain a large number of spectroscopic redshifts distributed across the sky.  This study alleviates these concerns by measuring
the angular overdensity between galaxies and quasars, resulting in a
much higher density of the tracer population.

We find no luminosity dependence of the quasar-galaxy cross-correlation function, consistent with previous findings.  A power law fit of luminosity to clustering amplitudes gives a slope of $-0.01 \pm 0.06$,
a much tighter constraint than those provided by previous studies at similar redshift.  We also fail to detect a relationship between clustering strength and MgII-based
virial black hole mass.  However, this result is consistent with theoretical predictions tightly linking black hole mass and host halo mass.
These results indicate that, within the ranges of luminosities considered here, the most luminous quasars reside in a wide range of dark matter halos.

\section{Acknowledgments}
\label{sec:ack}

We thank Doug Finkbeiner, Yue Shen, Adam Myers, Martin White, and Lars Hernquist
for helpful comments and suggestions.  We also thank Aaron Bray and Mario Juric
for assistance with the SDSS imaging mask and the selection queries for the
WISE and SDSS galaxies, respectively.  A.G.K. acknowledges support
from the Harvard College Research Program.

This publication makes use of data products from the Wide-field Infrared Survey Explorer, which is a joint project of the University of California, Los Angeles, and the Jet Propulsion Laboratory/California Institute of Technology, funded by the National Aeronautics and Space Administration.

Funding for SDSS-III has been provided by the Alfred P. Sloan Foundation, the Participating Institutions, the National Science Foundation, and the U.S. Department of Energy Office of Science. The SDSS-III web site is http://www.sdss3.org/.
SDSS-III is managed by the Astrophysical Research Consortium for the Participating Institutions of the SDSS-III Collaboration including the University of Arizona, the Brazilian Participation Group, Brookhaven National Laboratory, Carnegie Mellon University, University of Florida, the French Participation Group, the German Participation Group, Harvard University, the Instituto de Astrofisica de Canarias, the Michigan State/Notre Dame/JINA Participation Group, Johns Hopkins University, Lawrence Berkeley National Laboratory, Max Planck Institute for Astrophysics, Max Planck Institute for Extraterrestrial Physics, New Mexico State University, New York University, Ohio State University, Pennsylvania State University, University of Portsmouth, Princeton University, the Spanish Participation Group, University of Tokyo, University of Utah, Vanderbilt University, University of Virginia, University of Washington, and Yale University.

\bibliographystyle{apj}
\bibliography{final_report_refs_trunc}

\end{document}